\documentclass[sigconf,authorversion=true, nonacm=true]{acmart}
\usepackage{multirow}
\usepackage{enumitem}
\usepackage{subcaption}
\usepackage{bbm}
\usepackage{graphicx}
\usepackage{mathtools}
\usepackage[linesnumbered,ruled,vlined]{algorithm2e}
\usepackage{multirow}
\newtheorem{theorem}{Theorem}
\newcommand{\lpmname}{\textsf{GenNi}\xspace}
\newcommand{\saslpmname}{$\lpmname_{\text{SA}}$\xspace}
\newcommand{\ssslpmname}{$\lpmname_{S^3}$\xspace}

\AtBeginDocument{%
  \providecommand\BibTeX{{%
    \normalfont B\kern-0.5em{\scshape i\kern-0.25em b}\kern-0.8em\TeX}}}

\setcopyright{acmcopyright}
\copyrightyear{2021}
\acmYear{2021}
\acmDOI{nn.nnnn/nnnnnnn.nnnnnnn}

\acmConference[KDD 2022]{KDD 2022: The 28th ACM SIGKDD International Conference on Knowledge Discovery and Data Mining 2022}{August, 14-18, 2022}{Washington, DC, U.S.A}
\acmBooktitle{KDD '22: The 28th ACM SIGKDD International Conference on Knowledge Discovery and Data Mining 2022, August, 14-18, 2022, Washington, DC, U.S.A}
\acmPrice{15.00}
\acmISBN{XXX-X-XXXX-XXXX-X/YY/MM}

\setlist[itemize]{leftmargin=*}

\begin{document}

\title{
Generating Negative Samples for Sequential Recommendation
}

\author{Yongjun Chen$^{\clubsuit}$,
        Jia Li$^{\clubsuit}$,
        Zhiwei Liu$^{\clubsuit}$,
        Nitish Shirish Keskar$^{\clubsuit}$,\\
        Huan Wang$^{\clubsuit}$,
        Julian McAuley$^{\diamondsuit}$,
        Caiming Xiong$^{\clubsuit}$}
\affiliation{%
    $^{\clubsuit}$ Salesforce Research\\
    $^{\diamondsuit}$ UC San Diego\\
    \{yongjun.chen, jia.li, zhiweiliu, nkeskar, huan.wang, cxiong\}@salesforce.com, 
jmcauley@eng.ucsd.edu
}

\renewcommand{\shortauthors}{Y. Chen, et al.}

\begin{abstract}
To make \emph{Sequential Recommendation} (SR) successful, 
recent works
focus on
designing effective sequential encoders,
fusing side information, 
and mining extra positive self-supervision signals. 
The strategy of sampling 
negative
items at each time
step is less explored. 
Due to the dynamics of users' interests 
and model updates during training, 
considering randomly sampled items 
from a user's non-interacted item set 
as negatives can be uninformative. 
As a result, 
the model 
will
inaccurately learn user preferences toward items.
Identifying 
informative negatives
is
challenging
because informative negative items
are tied with both
dynamically changed interests and model parameters (and
sampling process should 
also be efficient).
To this end, 
we propose to \textbf{Gen}erate  \textbf{N}egative Samples (\textbf{i}tems) for SR
(\lpmname).
A negative item is sampled 
at each time step 
based on the current SR model's learned user preferences toward items.
An efficient implementation is proposed to further 
accelerate the generation process, making it scalable to large-scale recommendation
tasks.
Extensive experiments on four public datasets 
verify the importance of providing 
high-quality negative samples for SR and
demonstrate the effectiveness and efficiency of \lpmname.

\end{abstract}


\maketitle

\section{Introduction}

The central task of \emph{Sequential Recommendation} (SR) is to accurately
predict the next item that a user is interested in
based on her past behaviors (e.g., shopping, clicking, etc.).
To achieve this, 
an effective model must be able to 
learn accurate user preferences
toward massive vocabularies of items
at each time step.
Benefiting from the expressive power 
of deep neural networks (e.g., Transformer~\cite{vaswani2017attention,radford2019language}),
recent deep SR models including~\cite{kang2018self,li2020time,li2021lightweight,sun2019bert4rec,zhou2020s3,ma2020disentangled,qiu2021contrastive} 
arguably represent the current state-of-the-art.

Due to the high computational cost of computing the exact log likelihood for all items~\cite{zhou2021contrastive},
most SR methods are optimized
via a \emph{Noise Contrastive Estimation} (NCE)~\cite{gutmann2010noise,kang2018self,tang2018personalized,zhou2020s3} paradigm,
which is an approximation of maximum likelihood estimation~(MLE). 
Training with 
NCE requires the model to sample negative items to pair with positive items, where the training target is to
pull positive items closer to sequences while pushing away negative items.
Though existing methods improve SR from many different perspectives,
such as exploring the potential of different sequential encoders~\cite{hidasi2015session,tang2018personalized,tang2018personalized},
leveraging side information~\cite{li2020time,zhang2019feature,zhou2020s3}
and 
incorporating additional
training tasks~\cite{zhou2020s3,ma2020disentangled,liu2021contrastive,xie2020contrastive,qiu2021contrastive,qiu2021memory,chen2022intent,chen2022elecrec},
they rarely look into the impact of those negative items.
Instead, they commonly adopt uniform or popularity-biased sampling strategies, 
which are either unable to reflect true negative item distributions or sub-optimal for training sequence encoders.
Therefore, this paper 
investigates on the importance of 
sampling informative
negative items for training SR models.

\begin{figure}[t!]
  \centering
  \includegraphics[width=0.95\linewidth]{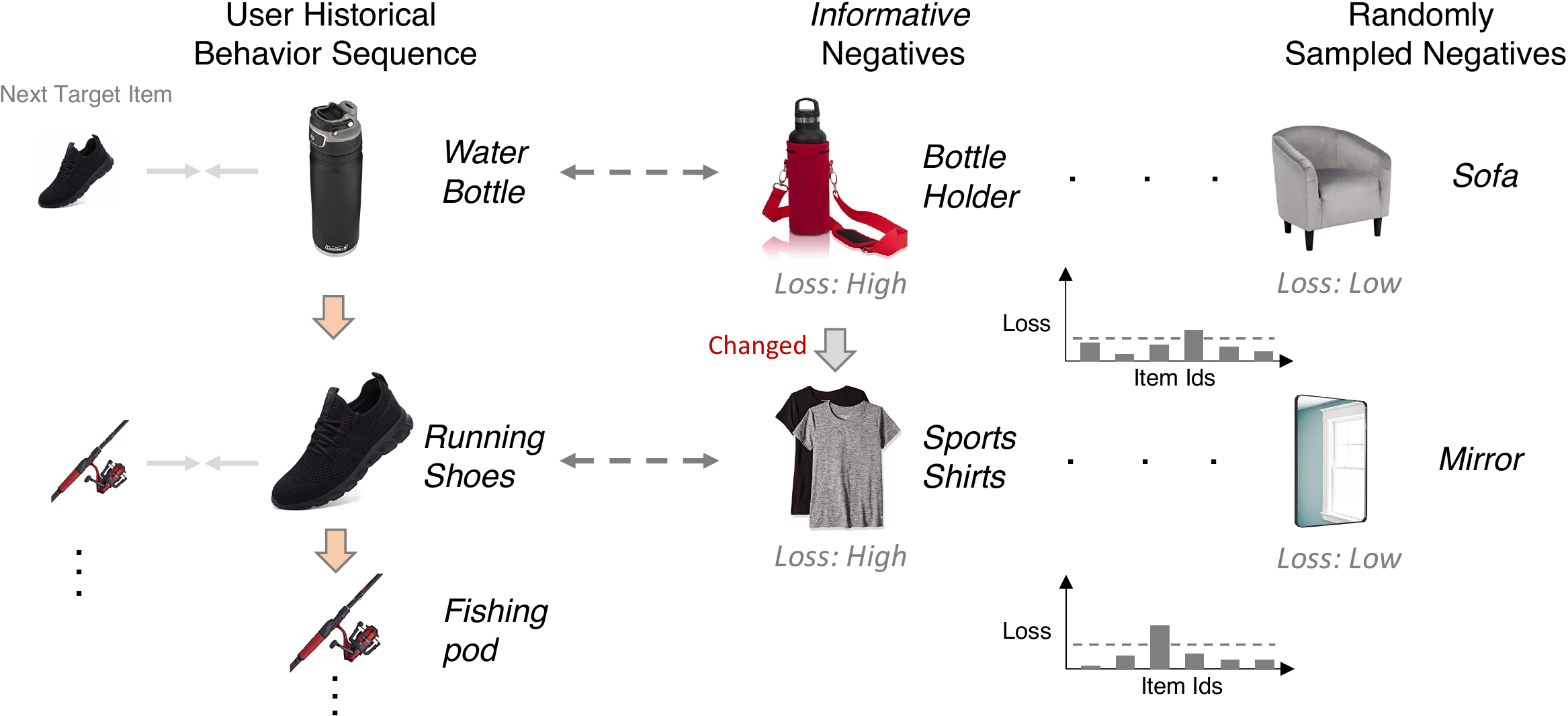}
  \caption{
  As a user’s interests change
  and as training goes on,
  the informative negative item distribution
  also changes.
  }
  \label{fig:motivation}
\end{figure}

To specifically demonstrate the necessity of sampling informative negative items in sequential recommendation, we illustrate a toy example in Figure~\ref{fig:motivation}.
When a user purchases a water bottle online,
the recommender predicts a bottle holder as her next 
item
because of 
observed concurrent consumption behavior from others;
however, she purchases shoes instead.
At this moment, the bottle holder is an informative negative because
the SR model made a wrong prediction.
After she purchases shoes, 
the informative negative item changes to sports shirts because
the model has observed sequential correlations between shoes and shirts, which however are not the actual next item in this user's sequence.
In such scenario, the uniform sampling method would ignore both dynamics and relatedness of negative items as training proceeds,
which samples uninformative items, 
thus contributing little to the optimization.  
Without 
dynamically generating informative 
negative samples, 
the SR model is unable to improve further,
resulting in sub-optimal performance of the sequential recommendation.

Nevertheless, sampling informative negative items for sequential recommendation
poses threefold challenges. 
First, it is non-trivial to characterize the sequential dynamics in 
sampling
informative negative items.
DNS~\cite{zhang2013optimizing} proposes a ranking-aware negative sampling scheme, which is devised to optimize static collaborative signals. 
PinSAGE~\cite{ying2018graph} identifies items of high PageRank scores with respect to positive items as the informative negative items.
MCNS~\cite{yang2020understanding} introduces a Markov Chain negative sampler for graph representation learning, which only harnesses constant neighbors for sampling. 
As in the aforementioned example,
the informative negative samples change according to users' consumption behaviors. 
In this sense, ignoring the sequential correlations fails to reveal true negative item distributions. 
Second, informative negative items are tightly associated with the model. 
During initial training stages, the model has no ability to classify items, so that all negative candidates are equally informative.
As training proceeds, the model is capable of identifying some negative items;
therefore, only those `hard' negative~\cite{bengio2008adaptive,blanc2018adaptive,robinson2020contrastive} items are informative and should be sampled to accelerate optimization.
As such, we should recognize the current state of models when generating negative samples.  
Last but not least, it is hard to retain efficiency. 
Due to the large-scale item corpus and (usually) sparse observed interactions, 
there are many negative item candidates.
Identifying informative negative items from those candidates requires awareness of their contributions to optimization, which is time-consuming. 
Therefore, it is crucial to efficiently
sample informative negative items to preserve the scalability of models.

To this end,
we propose to \textbf{Gen}erate  \textbf{N}egative \textbf{i}tems (\lpmname) for SR.
At each time step, 
a negative item is sampled 
based on 
the similarity between 
the current SR model learned user interests
and the item embeddings.
\lpmname adaptively generates negative samples 
without training an additional generative module except the SR model itself,
which reduces computation cost.
We develop an efficient algorithm
to further improve the computational efficiency
by a two-stage sampling strategy, 
which makes \lpmname scalable to
large-scale recommendation tasks.
A self-adjusted curriculum learning strategy is 
also proposed to
alleviate the human effort of tuning the hyperparameters in \lpmname. 
Though conceptually simple, \lpmname 
greatly improves upon state-of-the-art SR models.
Its success shares the same spirit with
works in other domains~\cite{kumar2010self,zhang2021understanding,zhang2013optimizing,sun2019rotate,yang2020understanding} 
where ``hard'' negatives
matter and can be generated 
via self-adversarial training.

We conduct extensive experiments on four public datasets
and observe that SR models
can be significantly improved simply by replacing
the original negative sampler with \lpmname,
e.g., the average performance of S$^3\text{-Rec}$~\cite{zhou2020s3} 
in NDCG@5 is improved $\textbf{107.66\%}$ over four datasets.
It shows that negative item sampling is as important as 
other components to make SR successful.
Detailed comparisons with other negative sampling strategies and analyses 
further validate the superiority of the proposed
method.

\section{Related Work}

\subsection{Sequential Recommendation}

Sequential recommendation aims
to accurately characterize users' dynamic interests
by modeling their past behavior sequences.
Early works on SR usually models
an
item-to-item transaction pattern
based on Markov
Chains~\cite{rendle2010factorization,he2016fusing}.
FPMC~\cite{rendle2010factorizing} 
combines the advantages of Markov Chains 
and matrix factorization 
to fuse both sequential patterns 
and users' general interest.
With the recent advances of deep learning,
many deep sequential recommendation models
are also developed~\cite{tang2018personalized,hidasi2015session,kang2018self,sun2019bert4rec}.
GRU4Rec~\cite{hidasi2015session}, 
Caser~\cite{tang2018personalized}, 
and SASRec~\cite{kang2018self} explore the potential of
encoding user sequential behaviors via an RNN, CNN, and Transformer,
respectively. 
FDSA~\cite{zhang2019feature}, TiSASRec~\cite{li2020time} and S$^3\text{-Rec}$~\cite{zhou2020s3} 
leverage side information (e.g., time-interval, item categories.)
for a comprehensive representation. 
BERT4Rec~\cite{sun2019bert4rec} 
replaces next-item prediction (NIP) task with a masked-item prediction task~\cite{taylor1953cloze} to capture contextual information.
With the success of contrastive self-supervised learning,
several works~\cite{ma2020disentangled,xie2020contrastive,qiu2021contrastive,qiu2021memory} propose different contrastive SSL paradigms
as a complement or a replacement task of NIP for a
more comprehensive learning. 
LSAN~\cite{li2021lightweight} also improves SASRec
from efficiency serving perspective 
Nevertheless, most existing works ignore the importance of 
quality of sampled negative items and view the item randomly 
sampled
from user non-interacted item set or all items in the
same training batch as negative items.

\subsection{Negative Sampling}
Word2vec~\cite{mikolov2013distributed}
first proposes to sample negative items
based on the word frequency distribution 
proportional to the 3/4 power
to train the skip-gram language models. 
Later works in NLP and Social Networks
often follow such 
setting~\cite{pennington2014glove,tang2015line,grover2016node2vec}.
In graph mining, RotatE~\cite{sun2019rotate} first
proposes to sample negative items
based on model's prediction
and then MCNS~\cite{yang2020understanding}
proposes to further improve its efficiency
via the Markov Chain based 
Metropolis-Hastings algorithm.
However, these methods only
consider neighborhoods of the nodes on graph
while ignore the sequential dynamic of the data.
Another line of works improves
the Sampled Softmax~\cite{zhou2021contrastive,blanc2018adaptive}
to better approximate to the full Softmax.
In contrast, our work 
study the SR methods that trained under NCE framework,
which trains a sequential binary classifier to distinguish 
target and negative items.
Several GAN-based~\cite{goodfellow2014generative} methods are proposed
for application such as information retrieval~\cite{wang2017irgan,park2019adversarial} 
and graph node
embeddings~\cite{cai2017kbgan}. 
However, GAN-based methods are often hard to train and
the additional training of generator also makes the 
sampling inefficient for SR models. In recommendation,
Bayesian Personalized Ranking~\cite{rendle2012bpr}
first proposes to sample negative items
uniformly from user non-interacted items
for training factorization machines.
Dynamic negative sampling(DNS)~\cite{zhang2013optimizing}
develops a ranking-aware negative sampling strategy
for improving collaborative filtering based methods.
PinSAGE~\cite{ying2018graph}
consider items with high PageRank scores as
``hard-negative'' samples
with curriculum learning
scheme to train large scale graph neural networks.
Despite of their success in their own domain, 
these methods ignored the importance
of the sequential dynamics
of users' interests thus are 
not ideal to be adopt for sequential recommendation.

\section{METHOD}

In this section, we  first describe 
the \emph{Sequential Recommendation} (SR) problem
and a general approach to solve the problem
with two key ingredients of training a SR model.
We then describe our proposed negative item generator,
an efficient algorithm as well as a self-adjusted 
curriculum learning approach to adaptively
sample negative items for each user.

\begin{figure*}[htb]
  \centering
  \includegraphics[width=0.95\linewidth]{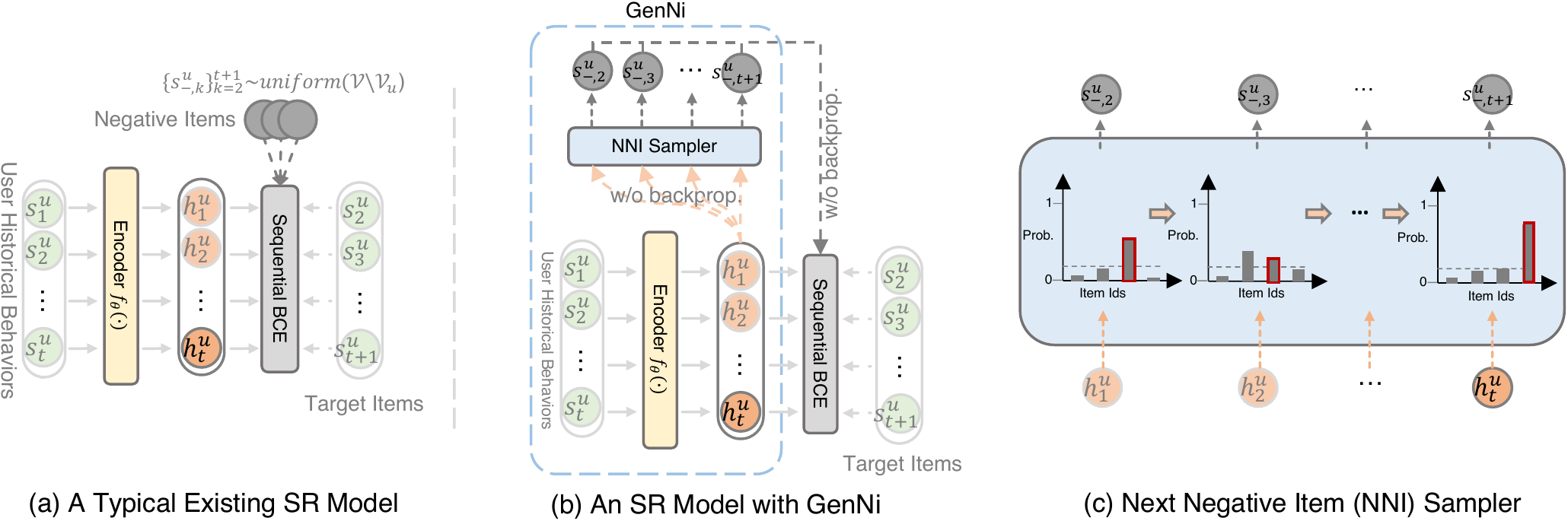}
  \caption{(a) is a typical sequential recommendation (SR) model.
  To learn the SR model, the next interacted item is viewed as a target item, 
  while a negative item is often sampled randomly 
  (e.g. uniformly) from the user non-interacted  
  item set at each time step.
  (b) Overview of an SR model with proposed \lpmname.
  (c) Illustration of how the Next Negative Item (NNI) Sampler works.
  At each time step, a negative item to a user is generated based on
  the user current interests. Intuitively, it picks an item that the model cannot
  identify it as a negative item more frequently for training.}
  \label{fig:model-illustration}
\end{figure*}

\subsection{Problem Formulation}
\label{sec:problem-definition}

SR is usually formulated as a \emph{next item prediction} (NIP) task.
Formally, in a recommender system, there is a set of users 
and items denoted as $\mathcal{U}$ and $\mathcal{V}$ respectively. 
Each user $u \in \mathcal{U}$ is associated with  a sequence of interacted items
sorted in chronological order $S^{u}= [s^{u}_{1}, \dots, s^{u}_{t}, \dots, s^{u}_{|S^{u}|}]$
where $|S^{u}|$ is the number of interacted items
and $s^{u}_{t}$ is the item $u$ interacted with at
step $t$. We denote $\mathbf{S}^{u}$ 
as the embedded representation of $S^{u}$,
where $\mathbf{s}^{u}_{t}$ is the $d$-dimensional embedding of item $s^{u}_{t}$. 
In practice, sequences are truncated with
maximum length $T$.
If the sequence length is larger than $T$, the most
recent $T$ actions are considered. If the sequence length is smaller than $T$, ``padding'' items will be 
added to the left until the length is
$T$~\cite{hidasi2015session,tang2018personalized,kang2018self}. 
For each user $u$ at time step $t$, 
the goal of SR is to predict
the item that the user $u$ 
would be interested in
at step $t+1$ among the item set $\mathcal{V}$,
given her past behavior sequence  $\mathbf{S}^{u}_{1:t}$.


\subsection{Training an SR Model with Noise Contrastive Estimation}
\label{sec:nce}

To train an SR model, a standard
learning procedure fits the sequential data following the maximum
likelihood estimation principle. Specifically,
for each user $u$ at position step $t$
in a mini-batch $\mathcal{B}$, 
we want to learn a parametric function $f_\theta$
that maximize the probability of the target item:
\begin{equation}
\label{eq:softmax-sum}
\arg \underset{\theta}{\max} \sum_{(u,t)\in B}P_{\theta}(\mathbf{s}_{t+1}^{u}|\mathbf{h}^{u}_{t}),
\end{equation}
where 
\begin{equation}
\label{eq:softmax}
P_{\theta}(\mathbf{s}_{t+1}^{u}|\mathbf{h}^{u}_{t}) = \frac{\exp(\mathbf{h}^{u}_{t}\cdot \mathbf{s}_{t+1}^{u})}{Z_{\theta}(\mathbf{h}^{u}_{t})},
\end{equation}
where $\mathbf{h}^{u}_{t}=f_{\theta}(S^{u}_{1:t})$ is the encoded user's interest representation at 
time $t$, $Z_{\theta}(\mathbf{h}^{u}_{t})=\sum_{v\in V}(\mathbf{h}^{u}_{t}\cdot \mathbf{v})$
is the  partition function
that normalizes the score into a probability distribution, and $\exp(\mathbf{h}^{u}_{t}\cdot \mathbf{s}^{u}_{t+1})$
is a similarity score of a user's preference toward the target item.
Unfortunately, computing this probability as well as 
its derivatives are infeasible 
since the $Z_{\theta}(\cdot)$ term requires 
summing over all items in $\mathcal{V}$, 
which is generally of large-scale in sequential recommendation.

Hence, existing methods~\cite{kang2018self,li2020time,ma2020disentangled,xie2020contrastive}  
commonly adopt an approximation 
via
\emph{Noise Contrastive Estimiation} (NCE)~\cite{gutmann2010noise}.
NCE is based on the reduction of density estimation to probabilistic binary classification.
It provides a stable and efficient way to avoid computing 
$Z_\theta(\cdot)$ while estimating the original goal.
The basic idea is to train a binary classifier to
discriminate between samples from the positive data distribution 
and samples from a ``noise'' (negative sampling) distribution.
Specifically,
given the encoded user interest $\mathbf{h}^{u}_{t}$, 
we view the next item $\mathbf{s}^{u}_{t+1}$
as its positive item and 
the sampled $k$
negative items from a pre-defined distribution function
$Q(\cdot)$ (e.g., a uniform distribution over all other items in $V$).
We train the SR model with the following loss function:
\begin{equation}
\label{eq:nce-all}
\begin{split}
\mathcal{L} = \sum_{(u,t)\in\mathcal{B}} \mathcal{L}_{t}^{u} 
\end{split}
\end{equation}
and 
\begin{equation}
\label{eq:nce}
\begin{split}
\mathcal{L}_{t}^{u} = & -\log(P(D=1|\mathbf{h}^{u}_{t}, \mathbf{s}^{u}_{t+1})) \\
& - k\mathbb{E}_{\text{neg}\sim Q}\log(P(D=0|\mathbf{h}^{u}_{t}, \mathbf{s}^{u}_{-,t+1})),
\end{split}
\end{equation}
where  $P(D=1|\mathbf{h}^{u}_{t}, \mathbf{s}^{u}_{t+1})=\sigma(\mathbf{h}^{u}_{t}\cdot \mathbf{s}^{u}_{t+1})$, $\sigma$ is sigmoid function, and $\mathbf{s}^{u}_{-,t+1}$ is the sampled
negative item at $t+1$.
This loss decreases when 
$\mathbf{h}^{u}_{t}\cdot \mathbf{s}^{u}_{t+1}$ increases and  
$\mathbf{h}^{u}_{t} \cdot \mathbf{s}^{u}_{-,t+1}$ decreases. 
In other words, optimizing this loss function is equivalent to pulling the sequence embedding
$\mathbf{h}^{u}_{t}$ closer to the positive item
$\mathbf{s}^{u}_{t+1}$
whilst pushing away from sampled negative items, thus being contrastive.
To make NCE approximate to maximum log-likelihood (Eq~\ref{eq:softmax}) closer, 
one needs to either sample more negative items 
or improve the quality of the negative sampling distribution $Q(\cdot)$. 
Surprisingly, neither of them is paid enough attention by existing methods.

\begin{theorem}[Impact of \(k\)]
\label{theorem-1}
Increasing \(k\) can reduce the mean square error (aka risk) of model
estimation and the distribution of negative items $Q(\cdot)$
become less important
when \(k \to \infty\).
\end{theorem}
The above theorem shows that $k$ is an importance factor
of making SR models well trained (proof given in Appendix~\ref{sec:the-1}).
Empirically, naively increasing
$k$ though trivial, but is not a good choice in recommendation tasks. 
Because under random sampling,
most of the sampled items can be uninformative 
with the training going on
while training time cost is 
linearly increased. 
Because of that, existing SR models often keep the default number $k=1$.
Without naively increasing 
the number of
negative items $k$, 
designing a good negative item distribution function $Q$
is crucial to make SR models successful.

\begin{theorem}[Optimal Embeddings]
\label{eq:opt-embs}
The optimal sequence and 
item embedding for each user $u$ at each time step $t$ should satisfy: 
\[ \mathbf{h}^{u}_{t}\cdot \mathbf{s}^{u}_{t+1} = 
- \log \frac{k \cdot Q(\mathbf{s}^{u}_{t+1}|\mathbf{h}^{u}_{t})}{P(\mathbf{s}^{u}_{t+1}|\mathbf{h}^{u}_{t})}. \]
\end{theorem}

Theorem~\ref{eq:opt-embs} indicates that 
the optimal embeddings are
dependent on both data distribution $P(\cdot)$ and the negative sampling distribution $Q(\cdot)$
(proof given in Appendix~\ref{sec:the-2}.).
As such, it is necessary to sample items from true negative sampling distribution, which would otherwise yield sub-optimal results.

The two theorems motivate us
to improve sampling process of negative items
for sequential recommendation as in following sections.
Hereafter, we propose a novel negative item generator 
as well as a strategy to further improve its efficiency.

\subsection{Next Negative Item Generator}
\label{sec:hn}

\subsubsection{\textbf{Principles of an Informative Negative Item Sampler in SR}}
\label{subsec:p-dis}

Theorem~\ref{eq:opt-embs} implies that in sequential recommendation, 
the informative negative items dynamically
change with the user's interests at time $t$
as well
as the network parameters $\theta$.
We therefore define the principles of informative 
negative item sampler for SR as follows:

\textit{Dynamic:} The sampler should be aware of the dynamic of 
the user's interests at each time step. 
When a user interacts with a new item, the corresponding 
informative negative items can also be changed.

\textit{Adaptive:} 
The sampler should be adaptive to the 
model structure as well as its parameters $f_{\theta}$.
The sampled item is uninformative 
if it is easy to be predicted as a negative item~\cite{kumar2010self}.

\textit{Efficient}: 
The sampler should also be efficient enough 
to scale to large recommender systems. 
The sampler can be alternated by tuning the hyperparameter $k$ or even
training without sampling (Eq.~\ref{eq:softmax-sum}) if it is inefficient.


\subsubsection{\textbf{Generating Negative Items via Self-Adversarial Training}}


Based on the aforementioned principles,
we propose to generate negative
items based on user's interests and model's 
current predictions.
Specifically, at each time step $t$, 
a user historical behavior sequence is encoded
by a networks: $\mathbf{h}_{t}^{u}=f_{\theta}(S_{1:t}^{u})$
(e.g., Transformer encoder~\cite{kang2018self,li2020time,ma2020disentangled,xie2020contrastive}).
Then we 
leverage the current sequential dynamic $\mathbf{h}_{t}^{u}$
and the model's current state (parameterized by $\theta$)
to generate \emph{next} informative negative item.
The $Q(\cdot)$ function is defined as follows:
\begin{equation}
\label{eq:q-function}
Q(s_{i}|\mathbf{h}_{t}^{u}, \hat{\theta}_{l}) = (\frac{\exp(\mathbf{s}_{i,\hat{\theta}_{l}} \cdot \mathbf{h}_{t,\hat{\theta}_{l}}^{u})}{\sum_{s_{i}\in \mathcal{V}} \exp(\mathbf{s}_{i,\hat{\theta}_{l}} \cdot \mathbf{h}_{t,\hat{\theta}_{l}}^{u})})^{\alpha}, s_{i}\neq s_{t+1}^{u},
\end{equation}
where $\hat{\theta}_{l}$ is the estimated model parameters at $l^{th}$ learning iterations
and $\alpha$ controls the difficulty of the sampler. When $\alpha=0$,
the sampler follows a uniform distribution.
The larger $\alpha$, the more informative item is more likely to be sampled.
We can see that now the $Q(\cdot)$ function is both \emph{dynamic} to the changes
of user's interests over each time step $t$ and also \emph{adaptive} to the model's learning state over each training iteration $l$.
We denote Eq~\ref{eq:q-function} next negative item (NNI) sampler.
The sampling strategy shares the same spirit with
works in other domains, such as CV, NLP and graph mining~\cite{bengio2008adaptive,kumar2010self,zhang2021understanding,zhang2013optimizing,sun2019rotate,yang2020understanding} 
where ``hard'' negatives
matter and can be generated via self-adversarial training.
Figure~\ref{fig:model-illustration} (b)-(c) illustrates
this process.

\begin{figure}[htb]
  \centering
  \includegraphics[width=0.85\linewidth]{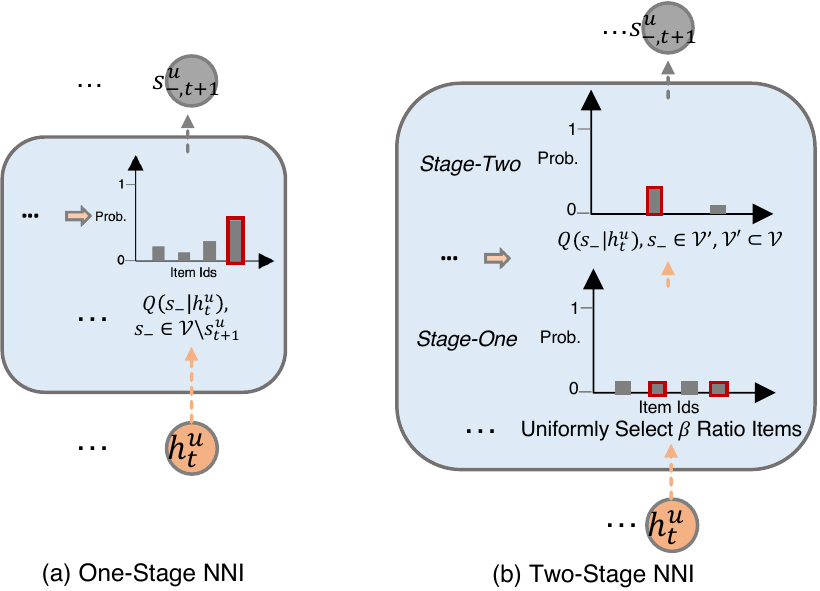}
  \caption{Comparison of the standard Next Negative Item (NNI) sampler and its acceleration. 
  }
  \label{fig:efficient-sampling}
\end{figure}

\subsubsection{\textbf{Acceleration}}
\label{sec:acc-gen}
Although Eq.~(\ref{eq:q-function}) already defines the negative sampling distribution, 
it is still inefficient due to the summation over all the items in the denominator part. Hence,
we devise a two-stage sampling strategy
to further accelerate the sampling procedure. 
To be more specific, 
at a certain time step,
a negative item is sampled as follows:
\begin{itemize}
    \item \textit{Pre-Selection:} 
    a small subset of candidate items 
    is pre-selected from $\mathcal{V}$
    in the first stage.
    We uniformly select $\beta$ 
    ratio of candidate items denoted 
    as $\mathcal{V'} 
  \subset \mathcal{V}$.
    \item\textit{Post-Selection:}
    we use the proposed NNI sampler
    to further narrow down the nominated
    items $\mathcal{V'}$ and serve to the user:
    \begin{equation}
    \label{eq:q-function-sampled}
    Q(s_{i}|\mathbf{h}_{t}^{u}) = (\frac{\exp(s_{i} \cdot \mathbf{h}_{t}^{u})}{\sum_{s_{i}\in \mathcal{V}'} \exp(s_{i} \cdot \mathbf{h}_{t}^{u})})^{\alpha}, s_{i}\neq s_{t+1}^{u},
    \end{equation}
\end{itemize}

With the acceleration, 
the computation time of negative item generation
reduces from the original $\mathcal{O}(|\mathcal{V}|)$ to $\mathcal{O}(\beta \cdot|\mathcal{V}|)$, 
where $\beta$ ranges from 0 to 1.
When $\beta \approx 0$,
sampling becomes uniform (and \textit{Post-Selection} is not needed).
When $\beta=1$, 
\textit{Pre-Selection} is no longer needed,
which becomes Eq.~\ref{eq:q-function}.
$\beta$ controls the trade-off between
effectiveness and efficiency. Figure~\ref{fig:efficient-sampling} illustrates the process. There are two strategies to set $\beta$:
\begin{itemize}
    \item \textbf{A fixed $\beta$ value.} 
    This strategy is simple and potentially can save the most computation cost.
    The drawback of having a fixed $\beta$ value is that
    as training proceeds, the number of informative items 
    become less and less (most of the items are already considered
    as negatives by the SR model). Having a small $\beta$ value
    can potentially filter out all the informative items in later training stage,
    so the model will stop learning. 
    Although, we empirically (in Section~\ref{sec:hpo}) find that $\beta$
    can be small without a large performance drop.
    \item \textbf{Gradually increasing $\beta$.}
    An alternative strategy is to gradually increase $\beta$
    as training proceeds:
    \begin{equation}
    \label{eq:dynamic-beta}
    \beta = \min (0.001 \cdot 10^{E_{i}/m}, 1.0),
    \end{equation}
    where $E_{i}$ denotes the $i^{th}$ training epoch and $m$
    controls how fast $\beta$ increases.
    Items sampled from a uniform distribution can be 
    informative in initial stages because the SR model hasn't started
    to learn. But most of them become uninformative as the training continues. By gradually increasing $\beta$,
    informative items can always 
    be sampled while reducing computation cost compared with 
    the full version (fixed $\beta=1.0$). 
\end{itemize}
See Section~\ref{sec:hpo} for more detailed comparisons.

\subsubsection{\textbf{Overall Scheme}}
We term the whole negative item generation 
process described from Section~\ref{subsec:p-dis} to Section~\ref{sec:acc-gen}
as \lpmname.
The 
overall training scheme with \lpmname for SR model is provided in
Algorithm~\ref{alg:genni-for-sr}.
It generates negative items based on the SR model
without introducing additional parameters.
The proposed acceleration strategy
further improves its efficiency so
\lpmname can be
scaled to large-scale recommendation tasks.
\lpmname is a model-agnostic negative item generator,
we apply \lpmname to both SASRec and  S$^3\text{-Rec}$,
denoted as \saslpmname and \ssslpmname.

\begin{algorithm}[htb]
\SetKwInput{KwInput}{Input}                
\SetKwInput{KwOutput}{Output} 
\KwInput{Users' historical behaviors 
$\{S^{u}_{1:T}\}_{u=1}^{|\mathcal{U}|}$,
sequential encoder $f_{\theta}$, hyper-parameters $\alpha$, $\beta$.}
\KwOutput{Learned $\theta$ including item embeddings $\{\mathbf{s}_{i}\}_{i=1}^{|\mathcal{V}|}$.}
 \While{ $epoch \leq MaxTrainEpoch$}{
    \For{a minibatch $\{S_{1:t}^{u}\}_{(u, t)\in \mathcal{B}}$}{
        \tcp{Sequential Encoding with \lpmname}
        \For{$(u, t) \in \mathcal{B}$}{
            \tcp{Encode Sequence via $f_{\theta}(\cdot)$} 
            $\mathbf{h}^{u}_{t} = f_{\theta}(S^{u}_{1:t})$ \\
            \tcp{Pre-selection with $\beta$ (fixed or gradually increasing)}
            $\mathcal{V}_{u}'= \text{Uniform}(\mathcal{V}, \beta)$ \\
            \tcp{Sample a Negative Item from $\mathcal{V}'$ via Eq.~\ref{eq:q-function-sampled}}
            $s^{u}_{-,t+1} \sim Q(\mathbf{h}_{t}^{u}, \mathcal{V}_{u}', \alpha)$ \\
            \tcp{View Next Item $s_{t+1}^{u}$ as Target Item}
        }
        \tcp{Next Item Prediction Optimization}
          Update $\theta$ based on 
          $\{\textbf{h}_{t}^{u}\}_{(u, t)\in \mathcal{B}}$,
          $\{s_{t+1}^{u}\}_{(u, t)\in \mathcal{B}}$,
          $\{s_{-,t+1}^{u}\}_{(u, t)\in \mathcal{B}}$
          to minimize the loss (Eq.~\ref{eq:nce}).
    }
 }
 \caption{\lpmname for Sequential Recommendation}
 \label{alg:genni-for-sr}
\end{algorithm}

\subsection{Self-Adjusted Curriculum Learning}
\label{sec:sa-cl}

\lpmname introduces $\alpha$
to control how often hard negatives are sampled.
But we must still manually tune
$\alpha$.
Curriculum learning~\cite{bengio2009curriculum,kumar2010self} 
allows neural networks 
to begin by understanding easy negative samples followed by hard ones.
We further reduce this rule to let the model 
itself adjust $\alpha$.
Specifically, we use the loss value in each batch as the critic 
to see if the current curriculum is too hard or too easy.
When the previous loss is larger than the current one,
we increase $\alpha$, otherwise we decrease $\alpha$.
In this way, $\alpha$ is self-adjusted with 
the online loss value as feedback, which
reduces human effort in choosing the initial 
$\alpha$ (see Section~\ref{sec:benfits-self-adjusted-cl}
for more detail).

\section{Discussion}

\subsection{Time Complexity and Convergence Analysis}
The computation costs of \saslpmname and \ssslpmname are 
similar to SASRec and  S$^3\text{-Rec}$
except that our methods use \lpmname instead of uniform sampling.
The overall computation cost is mainly from
Transformer, the feed-forward network and \lpmname,
which is
$\mathcal{O}(T^{2}\cdot d + T\cdot d^{2}
+ \beta \cdot |\mathcal{V}|\cdot T)$.
The dominant term is typically $\mathcal{O}(|T|^{2}d)$
from Transformer when $\beta$ is small.
Though \lpmname requires high computational cost when $\beta\cdot\mathcal{|V|}$ are large, 
however, our proposed acceleration strategy of it ensures faster convergence as well as better performance 
(see Section~\ref{sec:training-efficiency}).
The proposed two strategies of choosing $\beta$
in Section~\ref{sec:acc-gen} also help to balance the effectiveness and 
efficiency of \lpmname.
More details regarding convergence analysis are provided in Appendix~\ref{sec:convergence}.


\subsection{\lpmname for Improving Sequential BPR loss}
\label{sec:impro-bpr}
Though our method is induced from NCE paradigm in SR, 
\lpmname also has the ability to improve other training framework built upon pair-wise ranking loss, e.g.,
sequential BPR~\cite{rendle2012bpr}.
Previous work~\cite{hidasi2018recurrent} justifies that
optimizing a recommender model with a BPR loss results in gradient vanishing issue
if introducing more than one negative samples.
The reason is that after several epochs of training,
those uniformly sampled negative items already have lower scores than the target due to their easiness to identify. 
As a result, gradients towards those negative items gradually diminish. 
Instead, \lpmname generates informative negative items during each epoch of training,
which alleviates the gradient vanishing issue of BPR.
We conduct experiments to verify this claim in Section~\ref{subsec-other-optimizations}.

\section{Experiments}
\label{sec:experiments}

In this section, we evaluate the performance of our approaches
compared with the state-of-the-art sequential recommenders
and justify the benefits of our proposed negative item generator \lpmname.
We also investigate impacts of the hyper-parameters and 
conduct the ablation study. A case study is also 
included to better understand how \lpmname improves the training.

\begin{table*}[htb]
  \caption{Overall performance comparison among SR Models. 
  For each metric, the best score of our methods is in bold, and we underline the best scores in baselines.
  The last columns are the relative
  improvements compared between the bold and underlined scores.
}
  \label{tab:main-results}
  \setlength{\tabcolsep}{0.8mm}{
  \begin{tabular}{ll|ccccc|ccc|cc|c}
    \toprule
    \multicolumn{2}{c|}{\multirow{2}{*}{SR Model}} & 
    \multicolumn{1}{c}{\multirow{2}{*}{GRU4Rec}} & 
  \multicolumn{1}{c}{\multirow{2}{*}{Caser}}  &
    \multicolumn{1}{c}{\multirow{2}{*}{SASRec}}  &
    \multicolumn{1}{c}{\multirow{2}{*}{SASRec$_{\text{pop}_{\gamma}}$}} &
    \multicolumn{1}{c|}{\multirow{2}{*}{S$^3\text{-Rec}$}}   & 
    \multicolumn{1}{c}{\multirow{2}{*}{DSSRec}}  &
    \multicolumn{1}{c}{\multirow{2}{*}{CL4SRec}} &
    \multicolumn{1}{c|}{\multirow{2}{*}{MMInfoRec}} &
    \multicolumn{2}{c|}{\multirow{1}{*}{ours}} & 
     \multicolumn{1}{c}{\multirow{2}{*}{Improv. }}
      \\
      \cline{11-12}
      & &  & & & &  & &  & &
    \saslpmname&
    \ssslpmname& 
    \\

    \midrule
    \multirow{4}{*}{Beauty}  
    & HR@5   & 1.64 & 2.51 & 3.84$\pm$\small{0.06} & 4.08 & 3.85$\pm$\small{0.10} & 4.10 & 4.23$\pm$\small{0.31} &
    \underline{5.25}$\pm$\small{0.21} &
    6.30$\pm$\small{0.09} & \textbf{6.47}$\pm$\small{0.15} & 23.24\%\\
                             & HR@10  & 2.83 & 3.47 & 6.07$\pm$\small{0.11} & 6.18 & 6.35$\pm$\small{0.10}  & 6.89 & 6.94$\pm$\small{0.10} 
                             &
                             \underline{7.45}$\pm$\small{0.12} & 8.79$\pm$\small{0.05}& \textbf{9.45}$\pm$\small{0.21} &26.85\%\\
                             & NDCG@5 & 0.99 & 1.45 & 2.49$\pm$\small{0.09} & 2.69 & 2.40$\pm$\small{0.07} & 2.61 & 2.81$\pm$\small{0.18}
                             & \underline{3.71}$\pm$\small{0.06} & 4.48$\pm$\small{0.07} &  \textbf{4.64}$\pm$\small{0.04} & 25.07\%\\
                             & NDCG@10 & 1.37 & 1.76 & 3.21$\pm$\small{0.09} & 3.37 & 3.20$\pm$\small{0.07} & 3.58 & 3.73$\pm$\small{0.06} 
                             & \underline{4.43}$\pm$\small{0.10} & 5.33$\pm$\small{0.05} & \textbf{5.39}$\pm$\small{0.16} &  21.67\%\\
    \midrule
    \multirow{4}{*}{Sports}  
    & HR@5    & 1.62 & 1.54 & 2.20$\pm$\small{0.24} &2.22& 2.26$\pm$\small{0.03} & 2.14 & 2.17$\pm$\small{0.21} 
    &  \underline{2.78}$\pm$\small{0.09} & 3.55$\pm$\small{0.09}	& \textbf{3.68}$\pm$\small{0.13} & 32.37\% \\
	 & HR@10  & 2.04 & 1.94 & 3.41$\pm$\small{0.30} & 3.43& 3.73$\pm$\small{0.06} & 3.24 & 3.69$\pm$\small{0.09}  
	 & \underline{3.89}$\pm$\small{0.10} & 5.00$\pm$\small{0.11} 
	  & \textbf{5.50}$\pm$\small{0.09} & 49.05\%\\
	 & NDCG@5  & 1.03 & 1.14 & 1.45$\pm$\small{0.16} & 1.46&  1.45$\pm$\small{0.05} &1.42 & 1.37$\pm$\small{0.10}  
	 & \underline{1.91}$\pm$\small{0.08} & 2.57$\pm$\small{0.12} & \textbf{2.65}$\pm$\small{0.09} & 38.74\% \\
     & NDCG@10 & 1.10 & 1.424 & 1.84$\pm$\small{0.17}& 1.86 &1.93$\pm$\small{0.06}& 1.85 & 1.91$\pm$\small{0.08}  
     & \underline{2.33}$\pm$\small{0.11} & 3.04$\pm$\small{0.12} 
     & \textbf{3.14}$\pm$\small{0.08} & 34.76\%\\
    \midrule
    \multirow{4}{*}{Toys}  
    & HR@5  & 0.97 & 1.66 & 4.68$\pm$\small{0.16}   & 4.97& 4.43$\pm$\small{0.27} & 5.02& 5.26$\pm$\small{0.14}  
    &
    \underline{6.02}$\pm$\small{0.06} & \textbf{7.18}$\pm$\small{0.05}   
    & 6.96$\pm$\small{0.08} & 19.27\% \\
 	& HR@10  & 1.76 & 2.70 & 6.81$\pm$\small{0.19}  & 7.08& 7.00$\pm$\small{0.43} & 7.21& 7.76$\pm$\small{0.11} 
 	& \underline{8.14}$\pm$\small{0.08} &  \textbf{9.96}$\pm$\small{0.16}  & 9.50$\pm$\small{0.12} & 22.36\% \\
 	& NDCG@5 & 0.59 & 1.07 & 3.18$\pm$\small{0.09} & 3.37 & 2.94$\pm$\small{0.19} & 3.37 & 3.62$\pm$\small{0.08} 
 	& \underline{4.53}$\pm$\small{0.05} & \textbf{5.15}$\pm$\small{0.06}   
 	& 4.89$\pm$\small{0.08} &13.69\% \\
     & NDCG@10 & 0.84 & 1.41 & 3.87$\pm$\small{0.10} & 4.05 & 3.76$\pm$\small{0.24}& 4.21  & 4.28$\pm$\small{0.14}
     & \underline{5.10}$\pm$\small{0.04} &\textbf{5.90}$\pm$\small{0.05}  & 5.86$\pm$\small{0.09} &15.69\%\\
    \midrule
    \multirow{4}{*}{Yelp}  
    & HR@5    & 1.52 & 1.42 & 1.72$\pm$\small{0.04} & 1.73 & 1.94$\pm$\small{0.11} & 1.71 & 2.29$\pm$\small{0.03}
    & \underline{5.04}$\pm$\small{0.06} 
    & 5.25$\pm$\small{0.12}	&\textbf{5.35}$\pm$\small{0.02} & 6.15\%\\
     & HR@10  & 2.63 & 2.53 & 2.86$\pm$\small{0.03} & 2.88 & 3.35$\pm$\small{0.08} & 2.97 & 3.92$\pm$\small{0.10} 
     & \underline{6.01}$\pm$\small{0.09} & 7.72$\pm$\small{0.18}  &  \textbf{7.84}$\pm$\small{0.04} & 30.45\%\\
     & NDCG@5 & 0.91 &0.80 & 1.07$\pm$\small{0.03} & 0.99 & 1.19$\pm$\small{0.06} & 1.12 & 
     1.44$\pm$\small{0.01} 
     & \underline{3.19}$\pm$\small{0.08} & 3.28$\pm$\small{0.06} & \textbf{3.43}$\pm$\small{0.02} & 7.52\%\\
     & NDCG@10 & 1.34 & 1.29 & 1.44$\pm$\small{0.01} & 1.42& 1.64$\pm$\small{0.06} & 1.52 & 1.97$\pm$\small{0.05} 
     & \underline{3.60}$\pm$\small{0.13} & 4.03$\pm$\small{0.08} &  \textbf{4.15}$\pm$\small{0.01} &15.39\% \\
    
  \bottomrule
\end{tabular}}
\end{table*}

\subsection{Experimental Setup}

\subsubsection{\textbf{Datasets}}

We conduct experiments on four datasets: 
\emph{Sports}, \emph{Beauty}, \emph{Toys}, and \emph{Yelp}.
\emph{Sports}, \emph{Beauty}, and \emph{Toys}
are three subcategories 
of Amazon review data
introduced in~\cite{mcauley2015image}.
Yelp\footnote{https://www.yelp.com/dataset} 
is a dataset for business recommendation.
We follow~\cite{zhou2020s3,xie2020contrastive,ma2020disentangled,qiu2021memory} to prepare the datasets. In detail, we only keep the ``5-core'' datasets, in which all users and items have at least 5 interactions. 
The statistics of the prepared
datasets are summarized in Appnedix~\ref{sec:data-info}.

\subsubsection{\textbf{Evaluation Metrics}} 

For a fair comparison,  
we follow previous works~\cite{wang2019neural,krichene2020sampled} 
to rank the prediction on 
the whole item set without negative sampling. 
Performance is
evaluated on 
a variety of Top-K evaluation metrics, including \textit{Hit Ratio}$@k$ ($\mathrm{HR}@k$), and \textit{Normalized Discounted 
Cumulative Gain}$@k$ ($\mathrm{NDCG}@k$) where $k\in\{5, 10\}$.

\subsubsection{\textbf{Baselines}}
We compare our approach with three groups of representative baselines.
(i). SR models with uniform negative samplers
including
Caser~\cite{tang2018personalized}, GRU4Rec~\cite{hidasi2015session}, SASRec~\cite{kang2018self}, and S$^3\text{-Rec}$~\cite{zhou2020s3}.
We omit non-sequential models
such as 
BPR-MF~\cite{rendle2012bpr} and
simple item popularity based methods,
which are shown weaker than SR methods
on these datasets~\cite{sun2019bert4rec,zhou2020s3,ma2020disentangled}.
(ii). SR models with other negative sampling strategies
including DSSRec~\cite{ma2020disentangled}, CL4SRec~\cite{xie2020contrastive} and MMInfoRec~\cite{qiu2021memory}.
Different heuristic hard negative mining strategies are also purposed 
to further
improve the quality of negative samples.
(iii). Additional negative sampling strategies.
In addition, 
we also includes the popularity-based method~\cite{mikolov2013distributed} 
from NLP domain that
samples negative items based on the power
of item frequency $Q(i) \propto Pop(i)^{\gamma}$,
denoted as SASRec$_{\text{pop}_{\gamma}}$. 
Detailed descriptions of these baselines are in Appendix~\ref{sec-baseline-method}.

\subsubsection{\textbf{Implementation Details}}
Caser\footnote{https://github.com/graytowne/caser\_pytorch},
S$^3\text{-Rec}$\footnote{https://github.com/RUCAIBox/CIKM2020-S3Rec},
and 
MMInfoRec\footnote{https://github.com/RuihongQiu/MMInfoRec}
are provided by the authors.
GRU4Rec\footnote{{https://github.com/slientGe/Sequential\_Recommendation\_Tensorflow}}
and
DSSRec~\footnote{https://github.com/abinashsinha330/DSSRec}
are implemented based on public resources.
SASRec is implemented based on S$^3\text{-Rec}$
and we implement CL4SRec in Pytorch.
The number of attention heads and number of self-attention
layers in SASRec, S$^3\text{-Rec}$ and DSSRec
are tuned from $\{1, 2, 4\}$,
and $\{1, 2, 3\}$, respectively.
The number of latent factors introduced in DSSRec is tuned
from $\{1, 2,\dots, 8\}$.
For SASRec$_{\text{pop}_{\gamma}}$, 
we  tune the $\gamma$ from 0 to 1.5.

We implement two variants of our approaches
\saslpmname
and \ssslpmname
with Pytorch.
Our methods consider SASRec and S$^3\text{-Rec}$ as our base models
and replace the uniform sampler with our proposed \lpmname.
Models are optimized by an Adam optimizer~\cite{kingma2014adam}
with a
learning rate of 0.001, $\beta_{1}=0.9$, $\beta_{2}=0.999$, and
batch size of 256.
Early stopping criteria (models stop training if the performance on the validation set doesn't increase for 40 successive epochs) is used during training. 
For hyper-parameters in \lpmname, $\alpha$ is tuned from 0 to 6, 
$\beta$ is tuned from 0.0001 to 1.0 in a $\exp$ scale, number of negative items $k$
is tuned from 1 to 10.
We also provide results using a self-adjusted curriculum learning 
strategy (See Section~\ref{sec:sa-cl}) that reduces the need to tune $\alpha$.
All experiments are run
on a single Tesla V100 GPU
and 
we report the average results under 4 different random seeds on the test set.

All code shall be released upon publication.

\subsection{Performance Comparisons }
Table~\ref{tab:main-results} shows 
overall recommendation performance of all models
on the four datasets. We observe that:

\begin{itemize}
    \item Our methods \saslpmname and \ssslpmname both consistently outperform
    existing methods on all datasets by a large margin. 
    The average improvement
    compared with the best baseline ranges from \textbf{6.15\%} to \textbf{49.05\%}.
    Specifically, compared with SASRec and S$^3\text{-Rec}$,
    our approaches simply
    replacing their original uniform
    sampler with \lpmname, 
    achieve $\textbf{96.02\%}$ and $\textbf{107.66\%}$
    average 
    performance improvements on four datasets over SASRec and S$^3\text{-Rec}$
    at NDCG@5, respectively. 
    This observation clearly shows that sampling informative negative items
    is as important as other
    components in making SR successful and also demonstrates
    the effectiveness of our proposed sampler \lpmname.
    \item Transformer is an effective way of encoding
    user sequential dynamic patterns.
    Compared with GRU4Rec, Caser, SASRec and S$^3\text{-Rec}$
    we can see that  SASRec and S$^3\text{-Rec}$ that 
    utilize a Transformer-based encoder can consistently 
    achieve better performance compared to CNN/RNN-based encoders:
    Caser and GRU4Rec. 
    S$^3\text{-Rec}$ performs better than SASRec
    in most datasets because it fuses additional item attributes
    during pre-training. However, all these methods sample
    negative items randomly from user non-interacted item sets,
    yielding to a sub-optimally trained model.
    \item For different negative sampling strategies, SASRec$_{\text{pop}_{\gamma}}$ performs slightly
    better than SASRec, indicating that 
    the popularity-based method can help improve model learning.
    However, this strategy is static and does not consider the penalization
    of each user behavior, resulting in a large performance gap 
    compared to \saslpmname. 
    DSSRec, CL4SRec, and MMInfoRec
    proposes different contrastive self-supervised learning paradigms can 
    outperform other baselines that only train with an NIP objective.
    This observation demonstratives the effectiveness of
    contrastive self-supervised learning.
    These three methods commonly 
    consider items in the whole training batch as negatives, and
    MMInfoRec also proposes a heuristic hard negative mining strategy
    with a memory bank
    to further improve the quality of the samples.
    Their successes suggest that 
    sample more negative items and hard negative mining
    also benefits model learning. 
    Although MMInfoRec is the best baseline method, 
    it still performs worse than our approaches. 
    The reason might be twofold. First, 
    considering 
    all items in the training batch as negative items
    can introduce false-negative samples. 
    Second, heuristic hard negative mining (e.g., considering user historical interacted items as hard negatives in MMInfoRec) 
    is not adaptive
    to model parameters. As a result, 
    the sampled hard negatives can 
    gradually become uninformative to the model.

\end{itemize}

\subsection{Training Efficiency Comparison}
\label{sec:training-efficiency}

SASRec has
proven
to be an order of magnitude faster than CNN and RNN-based 
recommendation methods~\cite{kang2018self}, such as Caser and GRU4Rec.
In this section, we evaluate the efficiency of 
\saslpmname (on the Beauty dataset) by
comparing with the most efficient baseline SASRec and the best performing 
baseline MMInfoRec (See Appendix~\ref{app-nsc} for 
result comparisons on other datasets).
We omit the comparisons of \ssslpmname
as it has the same computation cost as \saslpmname
in its training stage. The only difference
is that \ssslpmname requires a pre-training stage to fuse
item attributes in the model.

\begin{figure}[htb]
  \centering
  \includegraphics[width=0.85\linewidth]{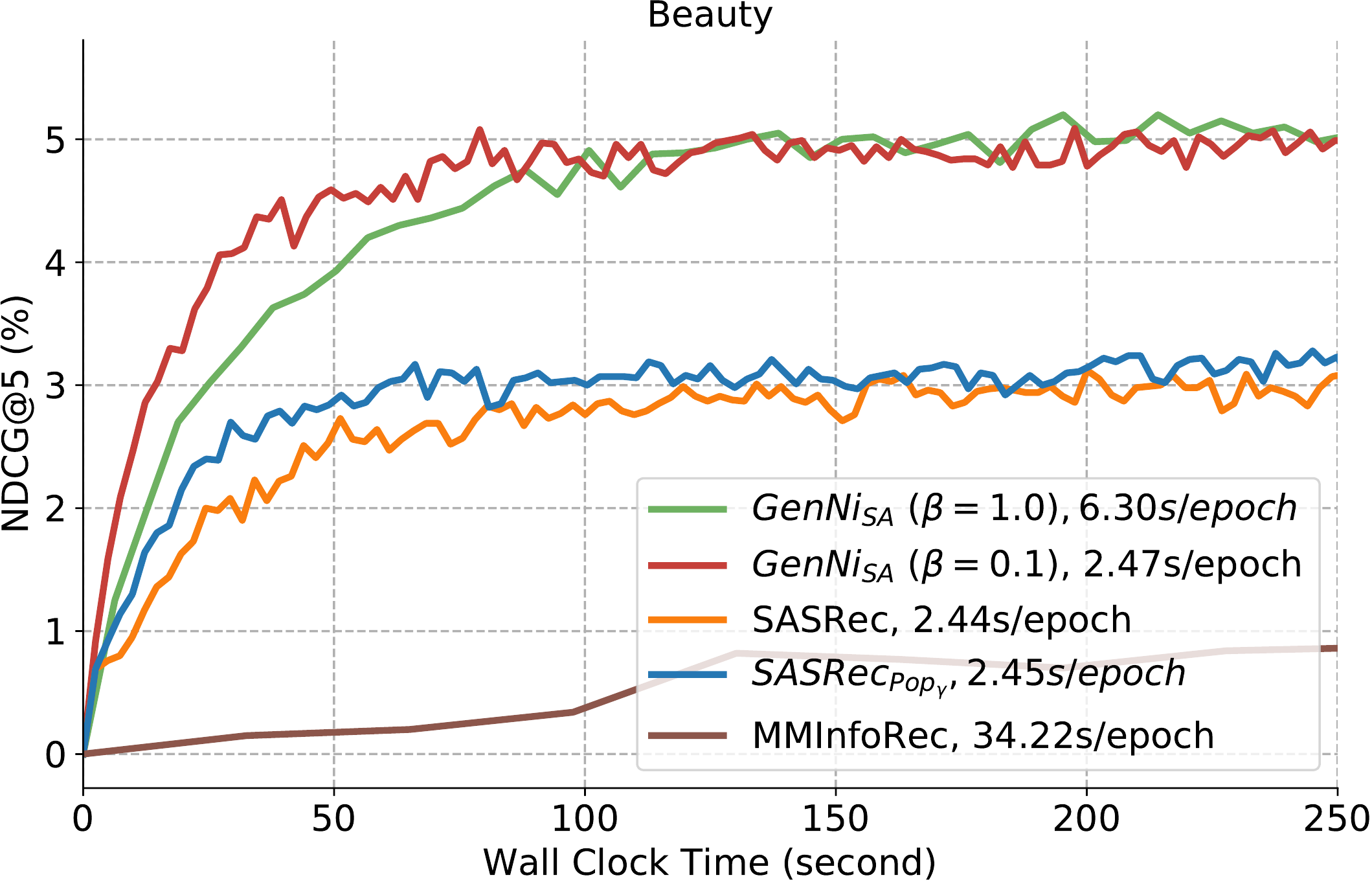}
  \caption{Validation Set Performance w.r.t.~Training time on
  the Beauty dataset. 
  }
  \label{fig:walk_clock_time}
\end{figure}

Figure~\ref{fig:walk_clock_time}
shows the performance w.r.t.~training (wall-clock) time
as well as the computation cost per epoch.
We can see that replacing the uniform sampler
with \lpmname does introduce additional computation cost;
for example,
SASRec spends 2.44 seconds on model
updates for one epoch while 
\saslpmname ($\beta=1$) requires 6.30s/epoch.
However, \saslpmname converges to much higher 
performance and requires fewer training epochs to converge.
What's more, as we reduce $\beta$ to 0.1,
\saslpmname ($\beta=0.1$) only needs
2.47 seconds to update the model for one epoch, which is
close to SASRec (2.44s/epoch), and still performs better
than SASRec.
Although MMInfoRec is the best performing baseline,
it requires 34.22 seconds on model updates for one epoch.
Our method \saslpmname ($\beta=1.0$) and \saslpmname ($\beta=0.1$) 
are over 5.42 and 13.85 times faster
and also perform better than MMInfoRec.

\subsection{Hyper-parameter Sensitivity}
\label{sec:hpo}
\lpmname introduces two hyper-parameters $\alpha$ and $\beta$
that controls the difficulty of sampled negatives 
and the negative item 
generation computation cost.
The number of negative samples is set as $k=1$,
which is the same as the original SASRec's setting for fair comparison.
We also study model sensitivity to the number of negative samples $k$,
the embedding size, and
learning rate.

\begin{figure}[htb]
  \centering
  \includegraphics[width=0.9\linewidth]{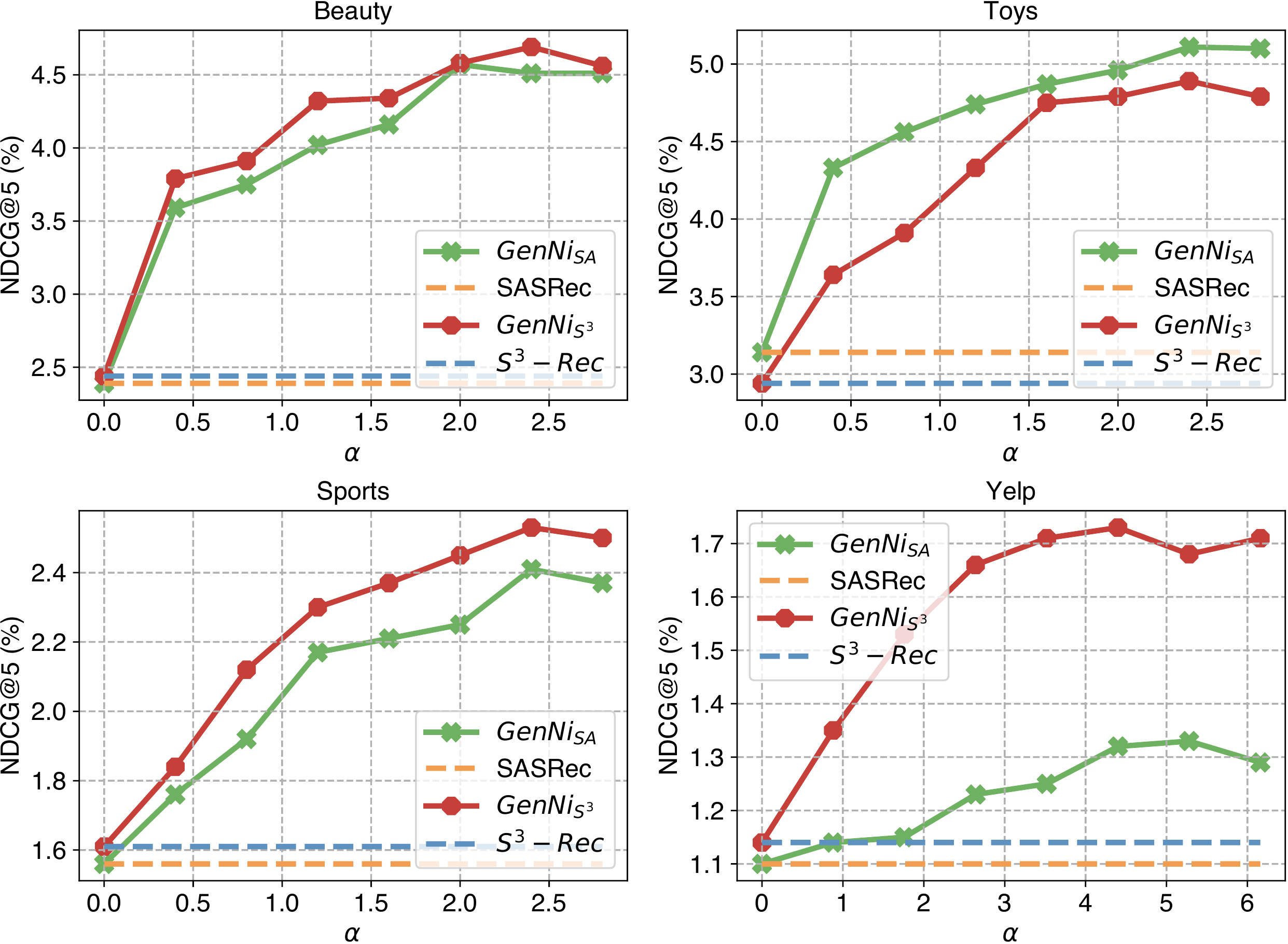}
  \caption{Performance w.r.t. $\alpha$ that controls the informativeness (difficulty) of sampled negative items. When $\alpha = 0$, negative items are uniformly sampled.} 
  \label{fig:alpha}
\end{figure}

\textbf{Impact of the informative of negative items $\alpha$.}
Figure~~\ref{fig:alpha} shows the influence of 
$\alpha$ on model performance over four datasets.  
We can see that the model performance increases as $\alpha$ increases 
at the beginning, and then the performance reaches a peak. Specifically, when $\alpha=2.5$, 
the model 
performs best on Beauty,while $\alpha=4.4$,
the model performs best on Yelp. 
Note that when $\alpha=0$, \lpmname becomes a uniform sampler.
The large $\alpha$ shows that
randomly sampled items can be uninformative as training proceeds, 
while considering items
that are currently hard to be correctly classified 
can further improve the model.
Similar observations are found on Sports and Toys. 

\begin{figure}[htb]
  \centering
  \includegraphics[width=0.9\linewidth]{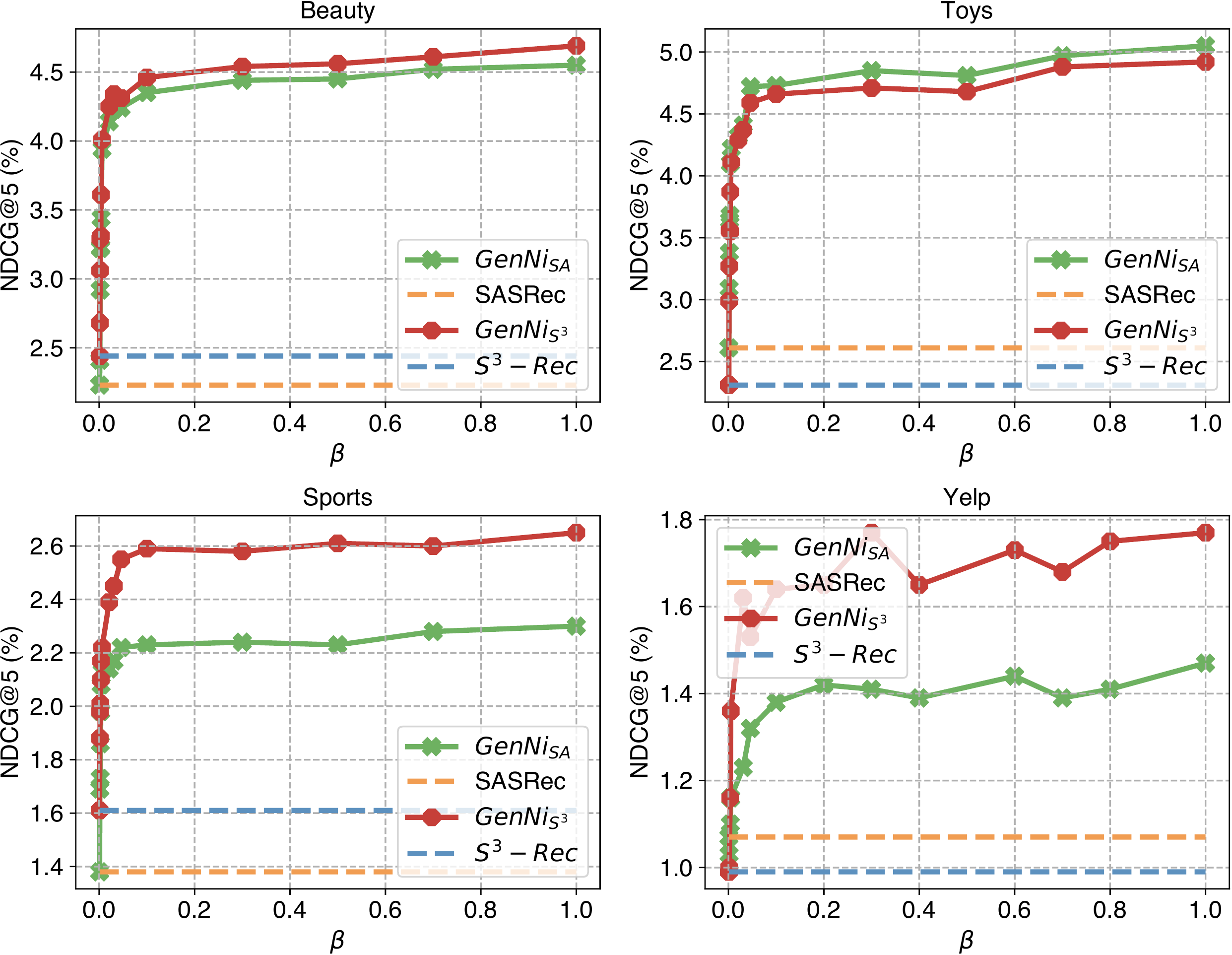}
  \caption{Performance w.r.t.~$\beta$ 
  for accelerating negative item generation.
  When $\beta \approx 0$, \lpmname is no longer needed and the negative items are sampled under a uniform distribution.}
  \label{fig:beta}
\end{figure}

\textbf{Impact of $\beta$ for accelerating generation.}
Figure~\ref{fig:beta} shows model performance w.r.t \textbf{a fixed $\beta$ value}.
We interestingly find that there is an elbow point of $\beta$
that balances the effectiveness and efficiency of \lpmname well.
For example, when $\beta=0.1$, it reduces about $90\%$ computation cost 
of \lpmname while the model can still achieve about \textbf{95\%} 
performance (e.g., NDCG@5) of its original version ($\beta=1.0$) in Beauty.
On one hand, it shows the superiority of 
\lpmname, which takes the efficiency of randomly sampling to pre-select 
a certain portion of items in the first stage and then
concentrates on finding informative ones with 
a slower but more accurate sampling strategy.
On the other hand, the decreasing of performance with small $\beta$
also indicates that with the training goes, the number of informative
items also decreasing so too small $\beta$ can filter out all these
items in pre-selection stage.
As introduced in Section~\ref{sec:acc-gen},
we also report the results that \textbf{gradually increasing
the $\beta$ value} via Eq~\ref{eq:dynamic-beta} in Table~\ref{tab:fix-or-flex-beta}. 
We can see that gradually increasing $\beta$ can achieves
the similar effect as of a fixed $\beta=1.0$ because the informative items
are decreasing along with training goes and $\beta$ can be small
while still capture informative items in early training stage.
This strategy reduces the computation cost while achieving same effect 
comparing with a fixed $\beta=1.0$.
\begin{table}[htb]
  \caption{Comparison of a fixed $\beta$ or gradually increasing
  $\beta$.}
  \label{tab:fix-or-flex-beta}
  \setlength{\tabcolsep}{1.0mm}{
  \begin{tabular}{c|c|cc|cc|cc}
    \toprule 
     \multicolumn{2}{c|}{\multirow{2}{*}{Strategy}} &
    \multicolumn{2}{c|}{Beauty} &
    \multicolumn{2}{c|}{Sports} &
    \multicolumn{2}{c}{Toys}  \\
    \cline{3-8}
    \multicolumn{2}{c|}{} & HR & NDCG & HR & NDCG & HR & NDCG\\
    \hline
    \multirow{2}{*}{\vtop{\hbox{\strut fixed $\beta$}}} & $\beta=0.1$  & 6.09 & 4.33 &3.18 & 2.14 & 6.50 & 4.72\\
    & $\beta=1.0$  & 6.30 & 4.48 &3.55 & 2.57 &7.18 & 5.15\\
    \hline
    \multirow{2}{*}{\vtop{\hbox{\strut Gradually}\hbox{\strut Increasing}}} & $m=20$ &6.35 & 4.53 & 3.50 & 2.52 & 7.16 & 5.07 \\
    & $m=40$ & 6.31 & 4.47 & 3.55 & 2.50 & 7.11 & 5.13\\

  \bottomrule
\end{tabular}}
\end{table}

\begin{figure}[htb]
  \centering
  \includegraphics[width=0.9\linewidth]{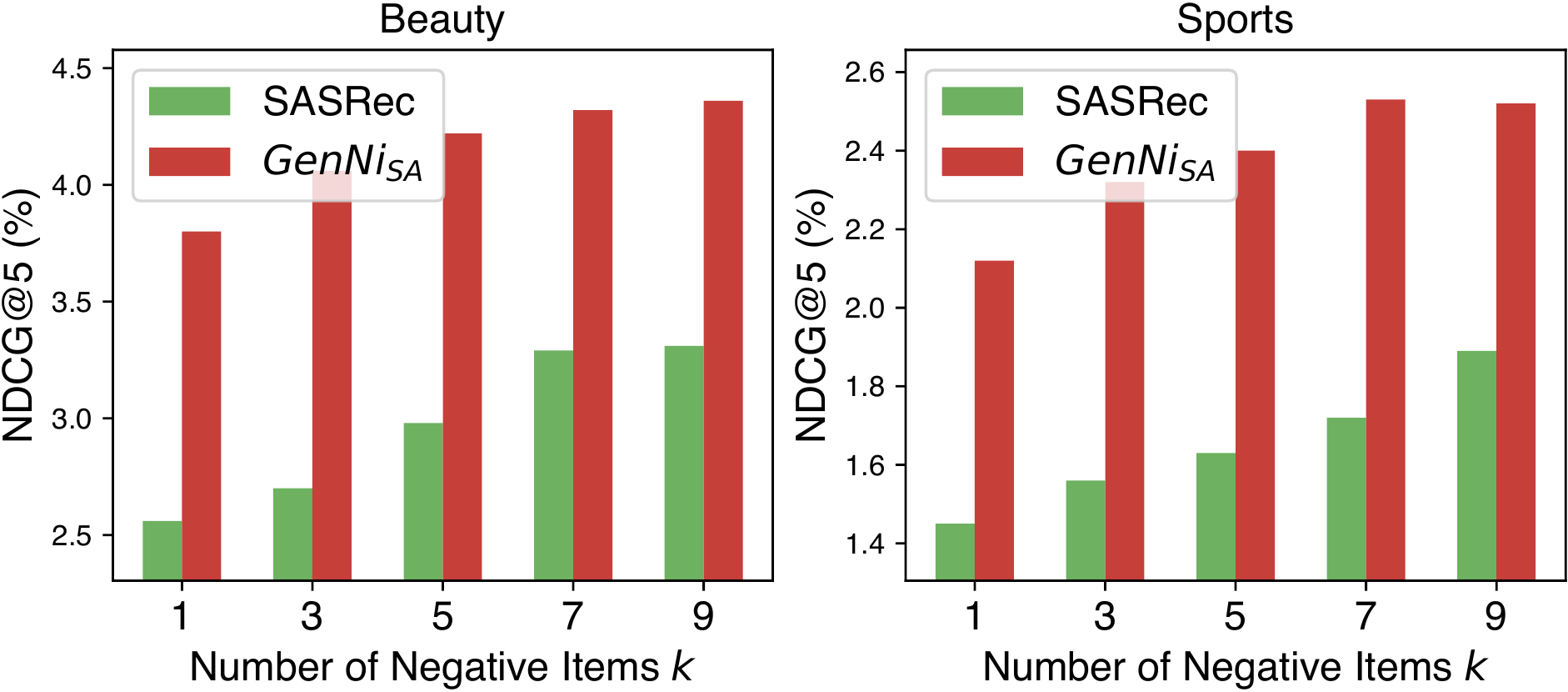}
  \caption{Performance w.r.t. of the number of negative items pairing
  with a target item on Beauty and Sports.}
  \label{fig:impact-of-k}
\end{figure}

\textbf{Impact of the number of negative samples $k$.}
Figure~\ref{fig:impact-of-k} shows the impact of the
number of negative samples. 
We can observe a diminishing return 
in the performance improvement
for both SASRec and \saslpmname.
However \saslpmname can consistently outperform
SASRec, which further verifies the importance of 
sampling informative negative items.
Note that training with additional negative samples
linearly increases the time cost.
While \saslpmname can even achieve better performance
with only 1 negative sample compared with 
SASRec that uses 9 negative samples on Beauty and Sports.
See Appendix~\ref{sec:additional-hpo} 
for additional results on Toys and Yelp,
and the sensitivity to 
the embedding size, and
learning rate.


\subsection{Ablation Study }

\subsubsection{\textbf{Benefits of Self-Adjusted Curriculum Learning}}
\label{sec:benfits-self-adjusted-cl}
\begin{figure}[htb]
  \centering
  \includegraphics[width=0.95\linewidth]{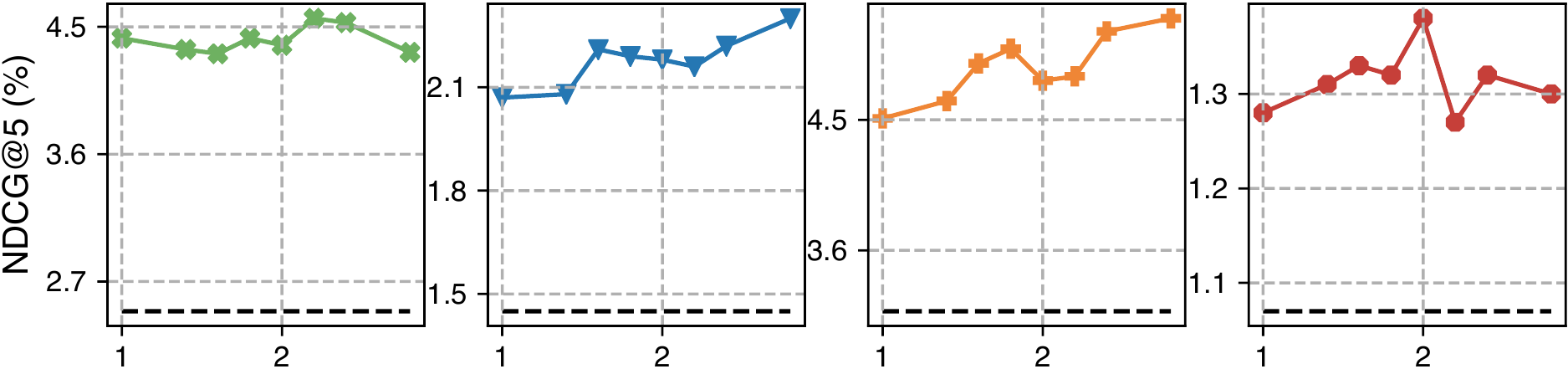}
  \caption{Performance w.r.t.~initial value of $\alpha$
  when employing self-adjusted curriculum learning on Beauty, Sports, Toys, and Yelp, respectively. The black color dash line is the performance of SASRec for comparison.}
  \label{fig:self-adj}
\end{figure}

As we can see from Figure~\ref{fig:alpha}, model performance is sensitive to $\alpha$;
in general, larger $\alpha$ benefits 
model performance.
In order to reduce the effort of tuning $\alpha$ for \lpmname,
we also propose a self-adjusted curriculum learning to let the model
adjust $\alpha$ based on its current performance. 
Figure~\ref{fig:self-adj} shows the sensitivity to the initial $\alpha$.
We can see the model performance is less sensitive to 
the initial $\alpha$ value.

\subsubsection{\textbf{\lpmname For Improving BPR Loss}}
\label{subsec-other-optimizations}
As discussed in Section~\ref{sec:impro-bpr},
training a SR model with sequential BPR loss can have a gradient vanish issue
when using additional negative samples ($k>1$).
In this section, we conduct experiments to show that \lpmname
can help alleviate such issues.
We train SASRec with a sequential BPR loss 
and replace the uniform sampling strategy 
used in BPR 
with \lpmname.
Table~\ref{tab:loss-compare} shows
comparisons
between uniform sampling and \lpmname in HR@5 and NDCG@5 (See Appendix~\ref{sec:additional-ablation-results} of additional results). 
We see that SASRec cannot benefit from more negative samples
when training with BPR loss because of the gradient vanishing issue.
After replacing the uniform sampler with \lpmname,
the model's 
performance
is improved with more negative samples.

\begin{table}[htb]
  \caption{Effectiveness of \lpmname for improving BPR loss (SASRec is the base SR model).}
  \label{tab:loss-compare}
  \setlength{\tabcolsep}{0.8mm}{
  \begin{tabular}{cc|cccccccc}
    \toprule 
    \multicolumn{2}{c|}{Additional Negatives} & 1 &
    2 &
    3 & 4 & 5 & 6 & 7 & 8 \\
    \hline
    \multirow{2}{*}{Uniform} & HR@5 & 2.32 & 2.16 & 2.21 & 2.34 & 2.13 & 2.14 & 2.24 & 2.07 \\
    & NDCG@5 & 1.42 & 1.28 & 1.33 & 1.36 & 1.27 & 1.31 & 1.34 & 1.25  \\
    \hline
    \multirow{2}{*}{\vtop{\hbox{\strut\lpmname}\hbox{\strut ($\alpha=2.2$)}}}& HR@5 & 5.64 & 5.7 & 5.83 & 5.81 & 5.93 & 5.87 & 5.96 & 6.08 \\
    & NDCG@5  & 3.90 & 4.04 & 4.11 & 4.07 & 4.12 & 4.16 & 4.23 & 4.25 \\
  \bottomrule
\end{tabular}}
\end{table}


\subsection{Case Study}
\begin{figure}[ht]
  \centering
  \includegraphics[width=1.0\linewidth]{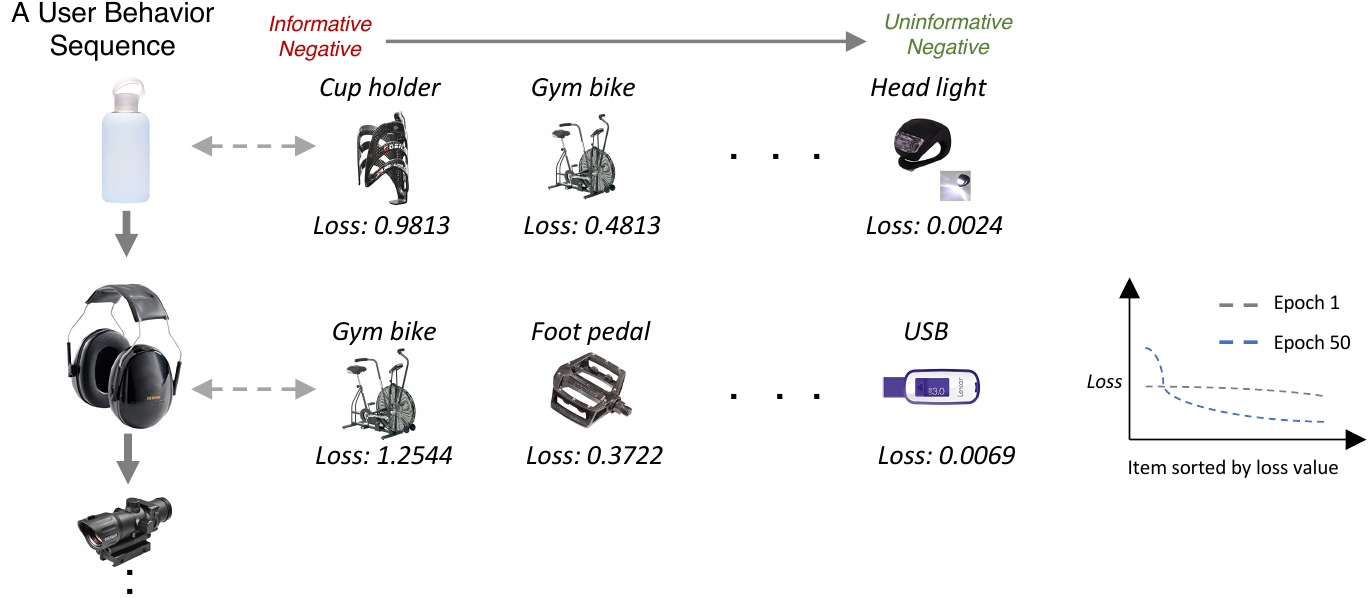}
  \caption{
  Visualization of dynamically changed informative negative items to model on the Sports dataset.
  }
  \label{fig:case_study}
\end{figure}

We conduct a case study on the Sports dataset~\cite{mcauley2015image}
to show examples of dynamically changing informative negative items.
Figure~\ref{fig:case_study} visualizes 
the informative items to the SR model.
When the user reviews a water bottle,
the cup holder is the most informative item;
the user reviews earphones instead,
and
the most informative items changes to a gym bike (etc.).
We can also observe 
that the informative negative distribution 
is close to uniform initially,
and gradually diversifies 
as training goes. 

\section{Conclusion}

In this work, we identified the dynamic of informative negative items
in sequential recommender systems,
because of the dynamic of users' interests, and the updates of model's
parameters
during training. 
We propose a negative item generator
\lpmname to adaptively generative informative negative samples
for training sequential recommenders.
Extensive studies on four datasets shows that
informative negative sampling is crucial of 
making the sequential recommendation model well-trained
and also demonstrates the superiority of \lpmname.
The detailed analysis also verified the effectiveness 
and efficiency of \lpmname.

\bibliographystyle{ACM-Reference-Format}
\bibliography{main}

\appendix


\section{Proof for Theorem~\ref{theorem-1}}
\label{sec:the-1}

We can derive from the discrete version of NCE theory (See~\cite{ma2018noise} 
for assumptions
that make the conclusion hold) 
that
there exists an integer $k_{0}$ such that, for a large sample size $|B|$,
for any $k>k_{0}$ (number of negative items),
\begin{equation}
\begin{split}
\label{conv}
\sqrt{\mathcal{|B|}|T|}(\hat{\theta}-\theta^{*})\Rightarrow \mathcal{N} (0, \mathcal{I}_{k}^{-1}),
\end{split}
\end{equation}

for some matrix $\mathcal{I}_{k}^{-1}$. So there exists
a constant value $C$ such that for any $k>k_{0}$,
the mean square error (MSE) (aka risk) of the parameter estimation is bounded by:

\begin{equation}
\begin{split}
\label{bound}
\mathbb{E}_{(u, t)\in \mathcal{B}}[||\hat{\theta}-\theta^{*}||^{2}] = & \frac{1}{|\mathcal{B}||T|}(\frac{1}{P(\mathbf{s}_{t}^{u}|\mathbf{h}_{t}^{u})} + \frac{1}{kQ(\mathbf{s}_{t}^{u}|\mathbf{h}_{t}^{u})}- \frac{k+1}{k}) \\
\leq & C / (k|T||\mathcal{B}|)
\end{split}
\end{equation}
As $k$ grows, the risk of parameter estimation is decreasing,
thus able to improve model performance. Alternatively, reader can 
follow~\cite{mnih2012fast} to calculate the gradient of Eq.~\ref{eq:nce} in terms of 
$\theta$ and will see that as $k \to \infty$, the gradient of 
Eq.~\ref{eq:nce} is approximated to the maximum likelihood gradient (Eq.~\ref{eq:softmax-sum}).
Eq.~\ref{bound} also shows that as $\mathcal{B}$ and $T$ larger and larger, $Q(\cdot)$
become less and less important as the estimation can be bound by $C/(k|T||\mathcal{B}|)$.
Interesting readers can read on~\cite{gutmann2012noise,ma2018noise} for a comprehensive review of 
NCE.

\section{Proof for Theorem~\ref{eq:opt-embs}}
\label{sec:the-2}
\begin{proof}

The SR model is optimized through the following objective:

\begin{equation}
\begin{split}
\label{overall-loss}
\mathcal{L} = \mathbb{E}_{(u, t)\in \mathcal{B}}\mathcal{L}_{t}^{u} = & -\mathbb{E}_{\mathbf{s}^{u}_{+,t+1}\sim P}\log(P(D=1|\mathbf{h}^{u}_{t}, \mathbf{s}^{u}_{+,t+1})) \\
   & - k\mathbb{E}_{\mathbf{s}^{u}_{-,t+1}\sim Q}\log(P(D=0|\mathbf{h}^{u}_{t}, \mathbf{s}^{u}_{-,t+1}))\\
 = & \sum_{t} [\sum_{\mathbf{s}_{+,t+1}^{u}}P(s_{+,t+1}^{u}|\mathbf{h}^{u}_{t})\log (\sigma (\mathbf{h}^{u}_{t} \cdot \mathbf{s}_{+,t+1}^{u})) \\
 + & k\sum_{\mathbf{s}_{-,t+1}^{u}}Q(s_{-,t+1}^{u}|\mathbf{h}^{u}_{t})\log (1- \sigma (\mathbf{h}^{u}_{t} \cdot \mathbf{s}_{-,t+1}^{u}))],
\end{split}
\end{equation}

where $\mathbf{s}^{u}_{+,t+1}$ and $\mathbf{s}^{u}_{-,t+1}$ are 
target and negative items to the user $u$ at time $t$. 
The above equation can be simplied as 

\begin{equation}
\begin{split}
\label{simplied-loss}
\mathcal{L} = & \sum_{t} \sum_{u} (P(s_{t+1}^{u}|\mathbf{h}^{u}_{t})+ K Q(s_{t+1}^{u}|\mathbf{h}^{u}_{t}))H(P', P'')
\end{split}
\end{equation}
where  $P'_{\mathbf{s}^{u}_{t+1},\mathbf{h}_{t}^{u}}(D=1)=\frac{P(\mathbf{s}^{u}_{t+1}|\mathbf{h}_{t}^{u})}{P(\mathbf{s}^{u}_{t+1}|\mathbf{h}_{t}^{u})+kQ(\mathbf{s}^{u}_{t+1}|\mathbf{h}_{t}^{u})}$
and $P''_{\mathbf{s}^{u}_{t+1},\mathbf{h}_{t}^{u}}(D=1)=\sigma(\mathbf{s}^{u}_{t+1}|\mathbf{h}_{t}^{u})$
are two Bernoulli distributions, and $H(\cdot)$ measures the cross entropy between two distributions.
Based on Gibbs inequality, optimized Eq~\ref{overall-loss} should satisfy that $P'=P''$ for all 
user interests $\mathbf{h}_{t}^{u}$ toward next predict item $\mathbf{s}^{u}_{t+1}$, i.e.,

\begin{equation}
\begin{split}
\label{otimal-ems}
\frac{1}{1+e ^{-\mathbf{h}^{u}_{t} \cdot \mathbf{s}_{-,t+1}^{u}}}= \frac{P(\mathbf{s}^{u}_{t+1}|\mathbf{h}_{t}^{u})}{P(\mathbf{s}^{u}_{t+1}|\mathbf{h}_{t}^{u})+k \cdot Q(\mathbf{s}^{u}_{t+1}|\mathbf{h}^{u}_{t})}.
\end{split}
\end{equation}
So the optimal embeddings should satisfy:
\begin{equation}
\begin{split}
\label{otimal-ems-2}
 \mathbf{h}^{u}_{t}\cdot \mathbf{s}^{u}_{t+1} = 
- \log \frac{k \cdot Q(\mathbf{s}^{u}_{t+1}|\mathbf{h}^{u}_{t})}{P(\mathbf{s}^{u}_{t+1}|\mathbf{h}^{u}_{t})}.
\end{split}
\end{equation}

\end{proof}

\section{Convergence Analysis}
\label{sec:convergence}

An explanation of why \lpmname is superior to heuristic samplings such 
as uniform sampler is that it can help
reduces the risk: $\mathbb{E}[||\hat{\theta}-\theta^{*}||^{2}]$.
From Eq~\ref{bound} we can see that,
as the training goes, the randomly sampled item would 
most likely has a small $Q(\mathbf{s}|\mathbf{h})$ than 
$P(\mathbf{s}|\mathbf{h})$ value, i.e.,
the model has learnt to classify it as a negative sample,
While the deviate in terms of $\theta$
is determined by the smallest value between
$Q(\mathbf{s}|\mathbf{h})$ and 
$P(\mathbf{s}|\mathbf{h})$. Optimize 
with small $Q(\mathbf{s}|\mathbf{h})$ in often time
interrupted the accurate optimization.
With \lpmname, the sampled negatives are
often has large $Q(\mathbf{s}|\mathbf{h})$ value
meaning that the estimation can more accurately
approximate to the optimal $\theta^{*}$.

\section{Data Information}
\label{sec:data-info}

The statistics of four datasets are shown in Table~\ref{tab:dataset-information}.
\begin{table}[htb]
  \caption{Dataset information.}
  \label{tab:dataset-information}
  \setlength{\tabcolsep}{1.6mm}{
  \begin{tabular}{c|cccccc}
    \toprule 
    Dataset   & Sports & Beauty & Toys & Yelp  \\
    \hline
    $|\mathcal{U}|$ & 35,598 & 22,363 & 19,412 & 30,431 \\
    $|\mathcal{V}|$ & 18,357 & 12,101 & 11,924 & 20,033  \\
    \# Actions      & 0.3m   & 0.2m  & 0.17m & 0.3m  \\
    Avg. length     & 8.3    & 8.9   & 8.6  & 8.3  \\
    Sparsity        & 99.95\%& 99.95\%& 99.93\%& 99.95\% \\
  \bottomrule
\end{tabular}}
\end{table}

\section{Baseline Methods}
\label{sec-baseline-method}
We compare our approach with three groups of representative baselines.

\begin{itemize}
    \item SR models with uniform negative samplers. GRU4Rec~\cite{hidasi2015session}, SASRec~\cite{kang2018self}, which encode sequences with CNN, RNN, and Transformer,
    respectively. S$^3\text{-Rec}$~\cite{zhou2020s3}, which  
    builds on SASRec with a pre-training stage to 
    incorporate additional item attributes into the model.
    We omit non-sequential models
    such as 
    BPR-MF~\cite{rendle2012bpr} and
    simple item popularity based methods,
    which are weaker than SR methods~\cite{sun2019bert4rec,zhou2020s3,ma2020disentangled}.
    \item SR models with other negative sampling strategies.
    We compare with recent works that 
    add or replace 
    the NIP objective with a contrastive self-supervised learning 
    objective: DSSRec~\cite{ma2020disentangled}, CL4SRec~\cite{xie2020contrastive} and MMInfoRec~\cite{qiu2021memory}.
    These works 
    follow the contrastive learning paradigm
    to 
    consider items in a training mini-batch as negatives 
    and propose 
    different heuristic hard negative mining strategies to further
    improve the quality of negative samples.
    respectively. 
    \item Additional negative sampling strategies. we also includes the popularity-based method~\cite{mikolov2013distributed} 
    from NLP domain that
    samples negative items based on the power
    of item frequency $Q(i) \propto Pop(i)^{\gamma}$,
    denoted as SASRec$_{\text{pop}_{\gamma}}$. 
\end{itemize}

\begin{table}[htb]
  \caption{Comparison of \saslpmname against other models (in HR@5) w.r.t the average (over 100 epochs) training time (second) per epoch.}
  \label{tab:ns-spped}
  \setlength{\tabcolsep}{1.0mm}{
  \begin{tabular}{c|c|cc|cc|cc|cc}
    \toprule 
     \multicolumn{2}{c|}{\multirow{1}{*}{Model}} &
    \multicolumn{2}{c|}{Beauty} &
    \multicolumn{2}{c|}{Sports} & \multicolumn{2}{c|}{Toys} &
    \multicolumn{2}{c}{Yelp}  \\
    \multicolumn{2}{c|}{} & time & HR & time & HR & time & HR & time & HR\\
    \hline
     \multicolumn{2}{c|}{SASRec} & 2.44 & 3.84 & 3.69 &2.20 & 2.09 & 4.68 & 3.35 & 1.72 \\
    \multicolumn{2}{c|}{SASRec$_{\text{pop}_{\gamma}}$} & 2.45& 4.08 & 3.66 & 2.12 & 2.11 & 4.97 & 3.36 & 1.58\\
    \hline
    \multicolumn{2}{c|}{MMInfoRec} &  34.22 &5.25 & 58.18 & 2.78 & 43.20 & 6.02 & 54.29 & 5.04 \\
    \hline
    \multirow{2}{*}{\vtop{\hbox{\strut\saslpmname}\hbox{\strut (vary $\beta$)}}} & 0.1 & 2.47 & 6.09 & 3.92 &3.18 & 2.17 & 6.50 & 3.39 & 2.08\\
    & 1.0 & 6.30 & 6.30 & 7.25 &3.55 & 3.13 &7.18& 6.56 &2.27 \\
  \bottomrule
\end{tabular}}
\end{table}

\begin{figure}[htb]
  \centering
  \includegraphics[width=0.85\linewidth]{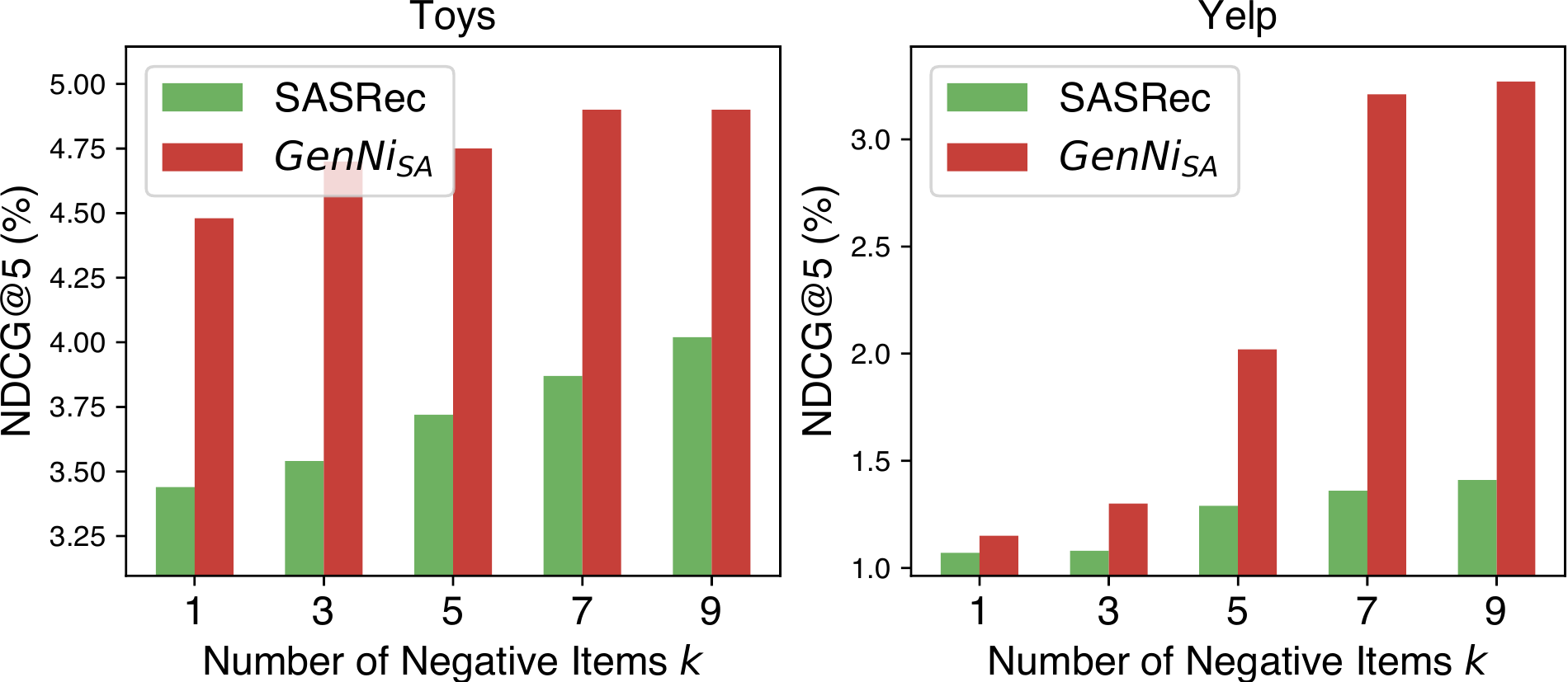}
  \caption{Performance w.r.t. of the number of negative items pairing
  with a target item on Toys and Yelp.}
  \label{fig:impact-of-k-additional}
\end{figure}

\section{Additional Efficiency Comparisons}
\label{app-nsc}

The training cost comparisons among SASRec, MMInfoRec, and \saslpmname over four datasets are reported in Table~\ref{tab:ns-spped}. In summary,
the $\beta$ of balances the effectiveness and efficiency of \lpmname.
Besides,
model with \lpmname (e.g., \saslpmname) can achieve better performance
than the best baseline MMInfoRec using much less computation resource.

\section{additional Results on Hyper-parameter Sensitivity}
\label{sec:additional-hpo}



\textbf{Impact of $k$}
Impact of $k$ on Toys and Yelp are shown in Figure~\ref{fig:impact-of-k-additional}.
Similar to observations on Beauty and Sports, 
Models can be further improved by sampling additional negatives
while \saslpmname can consistently outperform SASRec.

\begin{figure}[htb]
  \centering
  \includegraphics[width=0.85\linewidth]{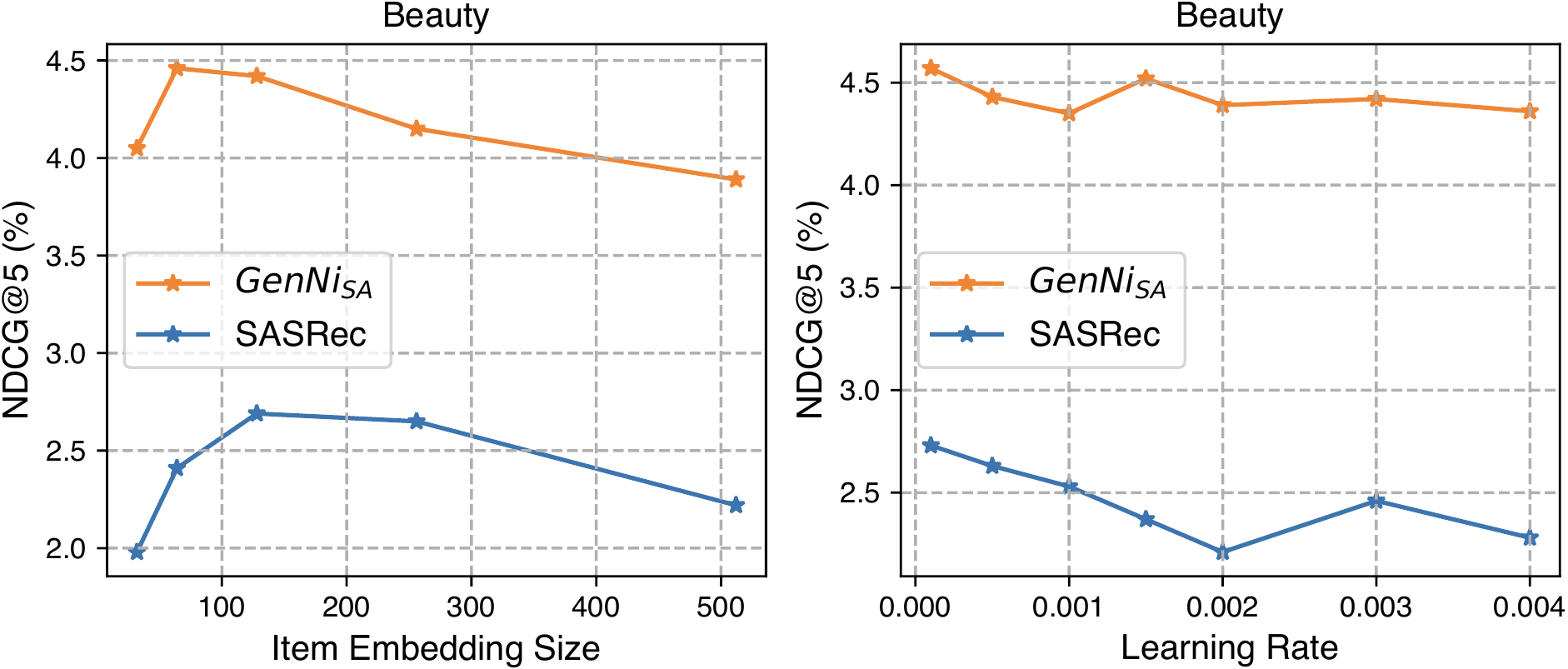}
  \caption{Performance of SASRec on Beauty varying the embedding dimensions and learning rate.}
  \label{fig:robust-to-model-hp}
\end{figure}

\textbf{Impact of embedding size and learning rate} Model's sensitivity to the embedding size and learning rate are shown in
Figure~\ref{fig:robust-to-model-hp}. We can see that vary learning rate 
or embedding size does influence model's final performance,
but their impact to SASRec and \saslpmname have a similar trend 
and \saslpmname can consistently outperform SASRec in all circumstances.

\begin{table}[ht!]
  \caption{Effectiveness of \lpmname for improving BPR loss (in HR@10 and NDCG@10).}
  \label{tab:loss-compare-additional}
  \setlength{\tabcolsep}{0.8mm}{
  \begin{tabular}{cc|cccccccc}
    \toprule 
    \multicolumn{2}{c|}{Additional Negatives} & 1 &
    2 &
    3 & 4 & 5 & 6 & 7 & 8 \\
    \hline
    \multirow{2}{*}{Uniform} 
    & HR@10 & 3.99 & 4.01 & 4.15 & 4.22 & 3.92 & 3.83 & 4.07& 3.89\\
    & NDCG@10 & 1.96 & 1.87 & 1.95 & 1.96 & 1.84 &1.86& 1.93 & 1.83 \\
    \hline
    \multirow{2}{*}{\vtop{\hbox{\strut\lpmname}\hbox{\strut ($\alpha=2.2$)}}}
    & HR@10 & 7.62 & 8.08 & 8.09 & 8.24 & 8.22 & 8.34 & 8.35 & 8.32 \\
    & NDCG@10  & 4.48 & 4.80 & 4.84 & 4.85 & 4.86 & 4.95 & 4.99 & 4.95 \\
  \bottomrule
\end{tabular}}
\end{table}

\section{Additional Results on Ablation Study}
\label{sec:additional-ablation-results}

Table~\ref{tab:loss-compare-additional} shows
the additional result comparisons between uniform
sampling and \lpmname in HR@10 and NDCG@10 with use of BPR loss.
We see observe that 
SASRec cannot benefit from more negative samples
when training with BPR loss. 
While \lpmname alleviates the gradient vanishing issue
thus the model's performance is stably improved 
after sampling more negative items.

\end{document}